\documentclass[12pt]{article}
\usepackage{amsfonts,amsmath,amssymb}
\usepackage[pdftex]{graphicx}
\usepackage{hyperref}
\usepackage{slashed}

\newcommand{\pa}{\partial}
\newcommand{\nn}{\nonumber}
\newcommand{\Ord}{{\cal O}}

\makeatletter

\@addtoreset{equation}{section}
\makeatother

\def\href#1#2{#2}
\begin{document}

\begin{titlepage}
\begin{center}

\hfill 
\vskip18mm

\textbf{\Large Weak Gravity Conjecture From}\\[3mm] 
\textbf{\Large Low Energy Observers' Perspective}\\[12mm] 

{\large Kazuyuki Furuuchi} 
\vskip6mm
\textit{Manipal Centre for Natural Sciences}\\
\textit{Centre of Excellence, Manipal Academy of Higher Education}\\
\textit{Dr.~T.M.A.~Pai Planetarium Building, Manipal 576104, Karnataka, India}

\vskip8mm 
\end{center}
\begin{abstract}
The Weak Gravity Conjecture (WGC) was proposed
to constrain
Effective Field Theories (EFTs)
with Abelian gauge symmetry
coupled to gravity.
In this article, I study the WGC from
low energy observers' perspective,
and revisit the issue of to what extent
the WGC actually constrains EFTs.
For this purpose, for a given EFT, 
I introduce
associated idealized low energy observers
who only have access to the energy scale
below the UV cut-off scale of the EFT.
In the framework of EFT, there is a clear difference
between the particles lighter than the UV cut-off scale
and the particles which are heavier than the UV cut-off scale,
as the lighter particles can be created below the UV cut-off scale
while the heavier particles are not.
This difference implies that the knowledge of
the low energy observers
on the stable heavy particles can be limited,
as the availability of 
the stable heavy particles is
determined by the environment prepared by some UV theory
unknown to the low energy observers.
The limitation of the knowledge
of the low energy observers 
regarding the stable heavy particles
whose mass is above the UV cut-off scale
of the EFT
leads to the limitation of the WGC
for constraining EFTs.
To illustrate these points in an example,
I analyze a model
proposed by Saraswat \cite{Saraswat:2016eaz}
which respects the WGC at high energy,
but which may appear to violate the WGC
for the low energy observers.
Implications of the analysis
to the bottom-up model buildings using EFTs are 
discussed.
\end{abstract}

\end{titlepage}

\tableofcontents

\section{Introduction}\label{sec:Intro}

Since the vastness of the string landscape 
began to be revealed, 
criteria for efficiently separating out
Effective Field Theories (EFTs) which
cannot be derived from string theory 
\cite{Vafa:2005ui} had been called for.
One of the most extensively studied
such criteria proposed so far is the
Weak Gravity Conjecture (WGC)
\cite{ArkaniHamed:2006dz}.
It requires that in an EFT
with an unbroken Abelian\footnote{%
The extension of the WGC to non-Abelian gauge theories 
have been discussed
\cite{Banks:2006mm},
but in this work I focus on properties particular
to Abelian gauge theories.}
gauge symmetry coupled to gravity,
there exists a particle to which
the electric Coulomb force acts stronger than 
the Newtonian gravitational force
(electric WGC).
Thus the electric WGC in $4$D requires 
the existence of a particle which satisfies\footnote{%
Throughout this article 
I will be concerned only
with parametric relations,
and I use $\sim$
to indicate that I are suppressing 
$\Ord(1)$ or even $\Ord(8\pi)$ numerical factors.}
\begin{equation}
\frac{g q M_P}{m} \gtrsim 1 \, ,
\label{eq:eWGCi}
\end{equation}
where $q$ is the charge\footnote{%
When I just write as charge, I mean the electric charge.
For a magnetic charge, I will explicitly mention it.} 
of the particle and $m$
is its mass, $g$ is the gauge coupling constant
and $M_P := \sqrt{8\pi G}$ is the reduced 
Planck mass ($G$ is the Newton's constant).
A physical motivation behind this conjecture
is an expectation that extremal Black Holes (BHs)
should be kinematically allowed to decay.
By requiring the existence of a magnetic monopole
to which the magnetic Coulomb force
acts stronger than the Newtonian gravitational force 
(magnetic WGC), 
together with an estimate of 
the mass of the monopole,
one obtains a bound on the UV cut-off scale
$\Lambda$ of the EFT \cite{ArkaniHamed:2006dz}:
\begin{equation}
\Lambda \lesssim g M_P \, .
\label{eq:cutWGCi}
\end{equation}
One of the motivations of the WGC was to use
this bound to explain
why extra-natural inflation had been difficult to realize
in string theory \cite{ArkaniHamed:2003wu}.
As can be seen from its origin,
aspects of the WGC as constraints on EFTs
have been of great interests,
though the WGC has also been extensively examined
in string theory which is the leading candidate
for a UV complete theory including quantum gravity.
In this article, I revisit the issue of
to what extent the WGC constrains EFTs.

The applicability of any EFT is limited 
by a certain UV cut-off scale $\Lambda$ 
above which the description by the EFT
breaks down.\footnote{Without gravity,
a Quantum Field Theory (QFT) can be UV complete, 
which corresponds to $\Lambda$ being infinity.
With gravity, 
it is natural to assume that 
$\Lambda$ is finite and below the Planck scale.
If one assumes asymptotically safe gravity,
it may accommodate infinite $\Lambda$,
but I will not consider such models.}
Some new physical degrees of freedom
must appear at $\Lambda$ to replace the description. 
In this framework of EFT,
there is a clear distinction 
between particles lighter than $\Lambda$ 
and those which are heavier: 
The particles whose mass is below the UV cut-off scale
of the EFT can be created within the EFT description,
while creations of heavy particles whose mass
is greater than $\Lambda$ 
cannot be described
within the EFT.
However, some
stable heavy particles created at the energy
above $\Lambda$ 
may 
remain at low energy.
If these stable heavy particles are coupled to gauge fields,
their interaction with the gauge particles could be of interest
and they may be incorporated in EFT.
Indeed, an EFT framework for describing
such stable heavy particles heavier than the
UV cut-off scale of the EFT had been developed in the past, 
one of the most important applications 
is in the study of heavy quark physics 
and known as Heavy Quark Effective Theory (HQET).\footnote{%
For a review of HQET, see for example 
\cite{Neubert:1996wg,Manohar:2000dt,Grozin:2004yc}.}
I will give a brief review of the
description of stable heavy particles in EFT
in Sec.~\ref{subsec:HPET}.

The point I would like to explore in this article
is that 
it is possible that
the low energy observers who only have access to
the energy below the UV cut-off scale $\Lambda$
may or may not know the existence of these stable heavy 
particles.\footnote{
The possibility that
charged particles which satisfy
the bound \eqref{eq:eWGCi}
all have mass above the UV cut-off scale of EFT
was
briefly mentioned in the original article 
\cite{ArkaniHamed:2006dz}.
However, the possibility that stable heavy particles
whose mass is beyond the UV cut-off scale 
could be known to the low energy observers
was not explored further, which I will do in the current article.}
The latter situation can happen if the stable heavy 
particles were not created with appreciable density,
or diluted away after they were created.
This is indeed a quite familiar setting
in particle cosmology, 
and
it could be of practical interests.
For example, one of the motivations of 
cosmic inflation
was to explain
why in our universe we have not observed
any magnetic monopole which was
predicted by Grand Unified Theories (GUTs)
\cite{Guth:1980zm,Einhorn:1980ik,Linde:1981mu}.
After all, one of the ultimate goals of string theory 
is to explain 
the particle physics realized in our universe,
and we have not observed
any such relics from high energy so far in our universe
(with possible exception of Dark Matters (DMs),
however our knowledge on their non-gravitational interactions
is very limited, we only have experimental bounds).
However, my interest here will be more on a matter of principle.
Thus, for a given EFT,
I will consider \textit{idealized} low energy observers
who can create and observe particles 
whose mass is below $\Lambda$. 
However,
I do \textit{not always} 
assume that the low energy observers
know all the stable heavy particles
whose mass is above the UV cut-off scale of the EFT,
since how much they have been created
is an input from the UV theory
and is not under control of the low energy observers.
This limitation of the knowledge of the 
low energy observers is a fundamental one
which cannot be improved by
the improvement of experimental techniques
(even if one has a perfect detector,
if there is no stable heavy particle left,
it is impossible to detect it).\footnote{%
If the density of the stable heavy particles 
left is small but finite, they can become detectable
by the improvement of the sensitivity of the detector.
However, such experimental practicalities are
not the interest of the current article.
In fact, what is relevant here is how much 
\textit{information} 
of the stable heavy particles 
the low energy observers have.
The environment prepared by the UV theory
and experimental ability both limit
this information, 
but the former is fundamental in a sense
that it cannot be changed
by the low energy observers.}
Stable heavy particles
might have been created in the early Universe or around
a small high-temperature BH.
However, these 
environments 
should be regarded
as a kind of natural particle colliders,
therefore 
if the observers are 
\textit{guaranteed} to have 
access to stable particles heavier than
the UV cut-off scale of the EFT 
produced from
these environments,
such observers should \textit{not} 
be regarded as the low energy observers
associated with the EFT.
Thus I analyze the cases one by one,
depending on
which stable heavy particles 
are known to the low energy observers.

There are at least two reasons to take into account
stable heavy particles when discussing the WGC.
The first reason is that 
it is of practical interest to ask
whether the WGC is satisfied
in the low energy IR EFT
when it arises from a UV EFT which satisfies the WGC.
If the WGC can be violated in the IR EFT, 
this means that
the constraining power of the WGC on low energy EFTs
can be quite limited.
However, if one assumes that the low energy observers
associated with IR EFT do not know any
particles whose mass is above the UV cut-off scale of IR EFT,
then when all the charged particles have mass above
the UV cut-off scale of the IR EFT,
the WGC \textit{appears} 
to be violated for the low energy observers
associated with the IR EFT.
For example,
take Quantum Electro-Dynamics (QED),
and consider
EFT obtained by
integrating out degrees of freedom 
above the energy scale of the electron mass.
There is no charged particle in the resulting EFT,
and the electric WGC appears to be violated in this case,
unless one considers the electrons whose mass is above
the UV cut-off scale of the EFT.
This example shows that
in order to obtain meaningful constraints on EFTs from the WGC,
one should take into account stable heavy particles
whose mass is above the UV cut-off scale of the EFT under consideration.
The second reason is that
once one accepts the magnetic WGC, in particular
the bound on the UV cut-off scale of EFTs
\eqref{eq:cutWGCi},
the electric WGC is automatically satisfied
by any charged particles whose mass is below the UV cut-off scale,
as will be explained in Sec.~\ref{subsec:trivial}.
This means that the electric WGC becomes 
a non-trivial question
only when there is no charged particle
whose mass is below the UV cut-off scale of the EFT under consideration.

The role of stable heavy particles
in constraining EFTs by the WGC
will be discussed in more detail in the main body.
To illustrate the ideas, 
I take
a simple model proposed by Saraswat \cite{Saraswat:2016eaz}
as an example.
This is an interesting model
which respects the WGC at high energy,
but which can violate 
a version of the electric WGC at low energy.
I further show that
depending on the knowledge of the low energy observers
on the stable heavy particles,
there are cases in which
the electric WGC  
\textit{appears} to be violated.
I also show that
the magnetic WGC is respected
in this model.
To show this result,
it is important to correctly identify 
the UV cut-off scale of the low energy EFT.
For this purpose,
I make detailed study of the internal structure of 
the monopole in this model which 
reveals the break-down of the low energy description.
The break-down of the low energy description
appears as fractional magnetic charges.
The fractional magnetic charges
would lead to observable Dirac strings
which would be problematic.
This is a break-down of the low energy description, 
and a new physics must enter 
to replace the low energy description.
The new physics which makes
the Dirac string unobservable here 
enters through
the Aharonov-Bohm effect,
which might have not been a 
common way for a new physics to appear
in the models studied in the past.

After the detailed analysis of 
the role of the stable heavy particles
in the Saraswat model,
I draw general lessons for
understanding actual constraints of the WGC on EFTs,
which is the main theme of this article.
Then, as an important example,
I take the extra-natural inflation model
based on the $5$D version of the
Saraswat model
and examine how the WGC constrains the model.

The organization of this article is as follows.
In Sec.~\ref{sec:EFT}, 
I review some key aspects
of the framework of EFT
relevant for discussing the WGC.
In particular,
I first review how to describe 
stable heavy particles
whose mass is above the UV cut-off scale of 
the EFT under consideration.
This framework is referred to as
Heavy Particle Effective Theory (HPET).
Then, using HPET, 
I give a precise statement
for what I mean by the low energy observers
associated with an EFT in this article.
The limitation of the knowledge 
of the low energy observers
associated with EFTs
regarding stable heavy particles
and its implications to the WGC are analyzed.
In Sec.~\ref{sec:srswt},
I take the model proposed by 
Saraswat to illustrate the points
discussed in the previous section.
Cases are classified
depending on which stable heavy 
particles are known to the low energy observers,
and are analyzed one by one.
From this study,
I draw general lessons for 
the bottom-up
model buildings using EFTs.
Then, the extra-natural inflation model
based on the Saraswat model
is studied in some detail.
In Sec.~\ref{sec:discussions},
I summarize the results and discuss
their implications towards 
the better understanding of the WGC.

\section{The WGC for low energy observers}\label{sec:EFT}

\subsection{Heavy Particle Effective Theory
(HPET)}\label{subsec:HPET}

I first review how to describe stable heavy particles 
whose mass is above the UV cut-off scale of an EFT,
in the framework of EFT.
The framework explained here had been applied 
in the study of heavy quark physics (Heavy Quark Effective Theory, HQET) 
(for a review of HQET,
see for example \cite{Neubert:1996wg,Manohar:2000dt,Grozin:2004yc}).
Below, I explain the framework in the case where
the stable heavy particle is a Dirac field,
but the framework can also be used for particles with 
different spins
\cite{Georgi:1990ak,Carone:1990pv}.

The starting idea for constructing the EFT
for stable heavy particles
is that in the infinite-mass limit,
the velocity of the heavy particles are conserved.
Therefore, the velocity is still a good quantum number
for particles with heavy but finite mass.
Thus one writes the four-momentum $p_\mu$ of a heavy particle state as 
\cite{Isgur:1989vq,Isgur:1989ed,Eichten:1989zv,Georgi:1990um}
\begin{equation}
p_\mu = m_h v_\mu + k_\mu \, ,
\end{equation}
where $m_h$ is the mass of the heavy particle and
the four-velocity $v_\mu$ is a time-like unit vector which satisfies
$v_0 > 0$ and $v \cdot v = 1$.
$k_\mu$ is the so-called residual momentum and 
is a measure of how much the heavy particle
is off-shell.
The low energy observers are those 
who are interested
in the low-energy processes which satisfy
\begin{equation}
| k_\mu | \ll \Lambda < m_h \, ,
\label{eq:LEA}
\end{equation}
where $\Lambda$ is the UV cut-off scale of the EFT.

The above idea is implemented at the action level as follows.
Let $\Psi$ be a Dirac field with mass $m_h > \Lambda$,
where $\Lambda$ is the UV cut-off scale of the EFT
with an Abelian gauge symmetry.
I start with the standard quadratic action for the Dirac field,
which corresponds to a UV theory for 
the EFT to be obtained:
\begin{equation}
S_{\Psi}
=
\int d^4x\,
\left[
\bar{\Psi} i {\slashed D} \Psi- m_h \bar{\Psi} \Psi 
\right] \,.
\label{eq:SPsi}
\end{equation}
Here, the covariant derivative is given as
\begin{equation}
D_\mu \Psi = 
\left(
\pa_\mu + i g q A_\mu
\right)
\Psi \, ,
\label{eq:covD}
\end{equation}
where 
$g$ is the gauge coupling,
$A_\mu$ is the Abelian gauge field
and $q$ is the charge of the field $\Psi$
with respect to the Abelian gauge group.

To obtain the low energy EFT appropriate for describing
processes with the residual momentum scale much lower than the 
UV cut-off scale (see \eqref{eq:LEA}),
one decomposes the field $\Psi$ 
into ``light part'' and ``heavy part'':
\begin{equation}
\Psi 
= 
e^{-i m_h v \cdot x} 
\left( h_v + H_v \right)
\, ,
\label{eq:psihH}
\end{equation}
where
\begin{equation}
h_v := P_+ \Psi \, ,
\quad
H_v := P_- \Psi \, ,
\quad
P_\pm : = \frac{1 \pm {\slashed v}}{2} \, .
\label{eq:hH}
\end{equation}
The field $h_v$ represents the light degrees of freedom
whereas
the field $H_v$ represents the heavy degrees of freedom,
as can be seen shortly.
Putting \eqref{eq:psihH} into
\eqref{eq:LEA} gives
\begin{equation}
S_{\Psi}
=
\int d^4x\,
\left[
\bar{h}_v 
i v \cdot D h_v  
-
\bar{H}_v
\left(
i v \cdot D
 +2 m_h
\right) H_v
+
\bar{h}_v i {\slashed D}_{\bot} H_v
+
\bar{H}_v i {\slashed D}_{\bot} h_v 
\right] \,.
\label{eq:ShH}
\end{equation}
Here, $D_{\bot \mu}$
is the covariant derivative in the direction
perpendicular to $v_\mu$:
\begin{equation}
D_{\bot \mu} : = 
\left(
\delta_{\mu}^{\nu} - v_\mu v^\nu
\right) D_{\nu} \, .
\label{eq:Dbot}
\end{equation}
From \eqref{eq:ShH},
one observes that $h_v$ is massless
whereas $H_v$ has mass $2m_h$ 
which is above the UV cut-off scale $\Lambda$.
By integrating out the heavy field $H_v$,
one obtains the effective action 
for the light degrees of freedom $h_v$.
Using the classical equation of motion:
\begin{equation}
\left( i v \cdot D + 2 m \right) H_v = i {\slashed D}_{\bot} h_v \, ,
\label{eq:ceq}
\end{equation}
one obtains the effective action for
the light field $h_v$:
\begin{equation}
S[h_v] :=
\int d^4x \,
 \left[
\bar{h}_v 
i v \cdot D h_v  
+
\bar{h}_v i {\slashed D}_{\bot} 
\frac{1}{2 m + i v\cdot D}
i {\slashed D}_{\bot} h_v 
\right] \, .
\label{eq:Sh}
\end{equation}
In the low energy regime \eqref{eq:LEA}
where the residual momentum is much smaller
than the mass of the heavy particle,
it would be appropriate to expand
the second term in
\eqref{eq:Sh} in Taylor series:
\begin{equation}
S[h_v] =
\int d^4x \,
 \left[
\bar{h}_v 
i v \cdot D h_v  
+
\frac{1}{2m_h}
\sum_{n=0}^\infty
\bar{h}_v i {\slashed D}_{\bot}
\left(
-\frac{i v\cdot D}{2m_h}
\right)^n
i {\slashed D}_{\bot} h_v 
\right]
\,.
\label{eq:Shl}
\end{equation}

In the above, the low energy effective action for the 
light degrees of freedom $h_v$ was derived
from the UV theory \eqref{eq:SPsi}.
On the other hand,
in the current article,
I will be interested in the case where
the low energy observers 
do not know the UV theory.
In this case,
the coefficients of the higher order terms in the 
effective action \eqref{eq:Shl}
are not fixed theoretically,
only experiments can fix them.

If one canonically quantize the action \eqref{eq:ShH},
$h_v$ corresponds to the field which
annihilates a particle with four-velocity $v$,
while
$H_v$ corresponds to the field
which annihilates an anti-particle with four-velocity $v$
(recall that I chose $v_0 > 0$).
Thus after integrating out $H_v$,
there is no degree of freedom which describes anti-particles.
To obtain an EFT which describes
stable heavy anti-particles, 
one replaces $v$ with $-v$
in the above procedures.

Below, I will refer to the EFT of stable heavy particles
explained in this subsection as
Heavy Particle Effective Theory, or HPET in brief.

\subsection{Low energy observers associated with an EFT 
and stable heavy particles}\label{subsec:heavy}

To study the WGC from low energy observers'
perspective,
for a given EFT,
I introduce a notion of 
low energy observers
associated with the EFT.
These low energy observers 
are assumed to have access to energy scale
below the UV cut-off scale $\Lambda$ of the EFT.
However, I do \textit{not always} assume
that these low energy observers
know \textit{all} the stable heavy particles
whose mass is above $\Lambda$. 
This is because these heavy particles
are not created below the energy scale $\Lambda$, 
thus their existence is controlled by
a UV theory which describes physics above $\Lambda$,
and it is not under the control of the low energy EFT.
	
The stable heavy particles whose mass is above 
the UV cut-off scale of the low energy EFT
and which are known to the low energy observers
should be included in the EFT in the framework of HPET, 
described in Sec.~\ref{subsec:HPET}.
Since I would like to consider idealized low energy observers
who can get all the information
contained in the EFT they are associated with,
technically they can be identified with the associated EFT itself.

To summarize,
what I mean by the low energy observers associated with an EFT
in this article is:
\begin{quotation}
EFT of fields whose mass is below the UV cut-off scale of the EFT,
and HPET of stable heavy particles whose mass is above 
the UV cut-off scale of the EFT.	
\end{quotation}

What I mean by 
``the stable heavy particles are known to the low energy observers'' is:
\begin{quotation}
The stable heavy particles are included in the EFT as HPET.
\end{quotation}

\subsection{Bound states in EFTs}\label{subsec:bound}

I briefly clarify my use of the terms
``particle states'' and ``bound states'' in EFTs.
In EFTs,
the internal structure of
a bound state whose size is below the UV cut-off (length) scale
$\sim 1/\Lambda$
cannot be resolved,
hence it is regarded as a point particle state in the EFT.
Therefore, by
bound state in EFT,
I mean that the state has a spatial structure 
whose size is bigger than $1/\Lambda$.
In this article, I will use the term particle state
exclusively for a point particle in EFTs, i.e. those state
which do not have an internal structure larger than $1/\Lambda$.
Thus in my definition,
bound states in EFT are not particle states.
These definitions justify my earlier statements 
that in the framework of EFT,
particles 
with mass larger than 
the UV cut-off scale of the EFT
$\Lambda$
cannot be created.
On the other hand, a bound state whose total energy 
is larger than $\Lambda$ can still be created,
since the constituent particles can be created
below $\Lambda$.

\subsection{The WGC for low energy observers}\label{subsec:verify}

Consider an EFT in $4$D with an unbroken
Abelian gauge symmetry which couples to gravity.
The 
electric WGC requires that 
there exists a particle whose 
charge $q$ and mass $m$ satisfies the bound
\begin{equation}
\frac{g q M_P}{m} \gtrsim 1 \, ,
\label{eq:eWGC}
\end{equation}
where 
$g$ is the coupling constant of the Abelian gauge theory
and
$M_P:=(8\pi G)^{-\frac{1}{2}}$
is the reduced Planck mass
($G$ is the Newton Constant).
The normalization of the
gauge coupling constant 
in Abelian gauge theories
depends on the convention,
even after one fixes the convention for
the product $g q$.
This point is not relevant for the electric WGC,
as the gauge coupling constant $g$ 
and the charge $q$ appears
in the combination $g q$.
However, it will be crucial when stating the magnetic WGC,
as I discuss shortly.
This is in contrast with non-Abelian gauge theories
in which the gauge field itself is charged
and one can use it as a standard for the normalization
of the gauge coupling constant.
One consequence is that 
the notion of the gauge coupling constant of an Abelian gauge theory
requires an existence of the minimal coupling to a charged particle.
In other words,
there is no physical meaning in
the gauge coupling constant of an Abelian gauge theory
if there is no charged particle in the theory.
In the context of the WGC,
the case in which there is no charged particle in the theory
may be regarded as a special case in which the electric WGC is violated
(this case is indeed a little bit special,
in that the EFT should have some
non-minimal coupling to the gauge field
in order for 
the low energy observers to probe the existence
of the gauge field).
I will use the convention in which the gauge coupling constant
of an Abelian gauge theory is normalized so that the 
smallest\footnote{Throughout this article,
when I state a charge is small or big,
what I am referring to is the absolute value of the charge.} 
charge is one.
However, 
here enters the limitation
of the low energy observers associated with
the EFT I explained in Sec.~\ref{subsec:heavy}:
The idealized low energy observers 
know all the charge and the mass of 
the particles whose mass is below the 
UV cut-off scale $\Lambda$ of the EFT,
but their knowledge
of the charge and the mass of 
the stable heavy particles whose mass
is above $\Lambda$ 
is limited
and depends on the environment prepared
by a UV theory.
This means that the low energy observers
may not know the existence of the particle
whose charge is truly the smallest.
Therefore, the more precise statement is that
in this article
\textit{I will normalize the gauge coupling constant 
of an Abelian gauge group in an EFT 
so that the smallest charge known to the 
low energy observers associated with the EFT is one}.
As discussed above,
making clear the convention used for the Abelian gauge coupling
is crucial when discussing the magnetic WGC,
but it has been implicit in earlier literature
and it seems to have become a source of confusions
in interpreting the Saraswat model,
which I will discuss in detail in Sec.~\ref{sec:srswt}.

There are several different versions 
of the electric WGC proposed in the past
\cite{ArkaniHamed:2006dz,%
Cheung:2014vva,%
Heidenreich:2015wga,Heidenreich:2015nta,%
Hebecker:2015zss,Heidenreich:2016aqi,%
Montero:2016tif}.
The difference is about
which type of particle satisfies the bound
\eqref{eq:eWGC}.
The weak version only requires that
a particle with the largest charge-to-mass ratio
satisfies the bound \eqref{eq:eWGC}.
The strong version requires that
the lightest charged particle
satisfies the bound.
The sub-lattice version 
requires that some $k \in  \mathbb{Z}$
and all $n \in \mathbb{Z}$,
there exist particles of charge
$q = k n$ with $g q M_P/m \gtrsim 1$.
Since this version requires a tower of charged states
whose masses are allowed to grow with charges,
it seems more appropriate to examine it 
in the top-down approach starting from
some fundamental theory like string theory.
Since my interest is on the constraints
on low energy EFTs from the WGC,
I will not discuss this version further in this article.
Another version, requiring that
a particle with the smallest charge
to satisfy the bound \eqref{eq:eWGC},
is known to have counterexamples
in string theory \cite{ArkaniHamed:2006dz}
and therefore
is not of further interest.

The magnetic WGC
requires that the
magnetic Coulomb force acts stronger to
a monopole with the smallest magnetic charge
than the Newtonian gravitational force:
\begin{equation}
\frac{g_m M_P}{m_{m 0}} \gtrsim 1 \, .
\label{eq:mWGC}
\end{equation}
Here, $m_{m0}$ is the mass of the monopole
with the smallest magnetic charge,
and
the magnetic gauge coupling constant $g_m$ is 
given in my convention
\begin{equation}
g_m = \frac{2\pi}{g} \, .
\label{eq:gm}
\end{equation}
The Dirac quantization condition
in this convention is
\begin{equation}
q \cdot q_m = n \, , \quad \mbox{($n$: integer),}
\label{eq:DiracQ}
\end{equation}
where $q$ is the electric charge and $q_m$ is the magnetic charge.
$q=1$ is the smallest electric charge 
known to the low energy observers, and
$q_m= 1$ is 
the smallest magnetic charge
\textit{still allowed}
for the given knowledge of the smallest electric charge
(more discussions on this point will be given in
Sec.~\ref{subsec:monopoles}).
Notice that
$q_m=1$ has been used in \eqref{eq:mWGC}.
The mass 
$m_{m0}$ 
of the monopole with the smallest magnetic charge (still allowed)
is estimated in the EFT as \cite{ArkaniHamed:2006dz}:
\begin{equation}
m_{m0} \sim \frac{\Lambda}{g^2} \, .
\label{eq:mmono}
\end{equation}
Putting \eqref{eq:gm} and \eqref{eq:mmono} into \eqref{eq:mWGC},
a bound on the UV cut-off scale $\Lambda$
of the EFT is obtained \cite{ArkaniHamed:2006dz}:
\begin{equation}
\Lambda \lesssim g M_P \, .
\label{eq:cutWGC}
\end{equation}

The bound \eqref{eq:cutWGC} can
also be obtained by requiring that
the monopole with 
the smallest magnetic 
charge is not a BH \cite{ArkaniHamed:2006dz}.
The size 
of the monopole 
with the smallest magnetic charge is
determined 
by the UV cut-off scale $\Lambda$ of an EFT.
Assuming that it is bigger than the Schwarzschild radius:
\begin{equation}
\frac{1}{\Lambda} \gtrsim \frac{m_{m0}}{M_P^2} \, ,
\label{eq:notBH}
\end{equation}
with the estimate of the mass $m_{m0}$ 
of the monopole with the smallest magnetic charge \eqref{eq:mmono},
one again obtains the bound on the UV cut-off scale $\Lambda$ of the EFT
\eqref{eq:cutWGC}.

It is important to
notice that 
while the physical requirements
which led to the bound on the UV cut-off
did not depend on the convention,
\textit{the expression \eqref{eq:cutWGC}
depends on the normalization convention 
of the gauge coupling constant}.
This was because 
the physical arguments which led to \eqref{eq:cutWGC}
involved the knowledge of the smallest magnetic charge
which was dependent on the knowledge of the electric charges, 
which in turn was needed for
a physically meaningful normalization 
of the gauge coupling constant.
Making clear the convention used in \eqref{eq:cutWGC} 
is crucial in models
in which there exists a large hierarchy between charges of particles.
The Saraswat model which
I will study in Sec.~\ref{sec:srswt}
is an example of such models.
As declared earlier, 
I use the convention in which
the smallest charge known to 
the low energy observers associated with
the EFT is normalized to one
to state \eqref{eq:cutWGC},
which has an advantage 
that the Dirac quantization condition
\eqref{eq:DiracQ} becomes simple.

\subsection{When the magnetic WGC is satisfied,
the electric WGC bound is
automatically satisfied by
charged particles which
are lighter than the UV cut-off scale of 
the EFT}\label{subsec:trivial}

After making clear the convention of the 
gauge coupling in the previous subsection,
I can now state an important fact:
{\em When the bound on the UV cut-off scale of the EFT
\eqref{eq:cutWGC}
holds,
any charged particle whose mass is below the 
UV cut-off scale of the EFT automatically satisfies the
electric WGC bound \eqref{eq:eWGC}.}
To see this,
let $q$ be the charge of a particle
whose mass $m$ is below the UV cut-off scale $\Lambda$,
i.e. $m < \Lambda$.
Then it follows that
\begin{equation}
\frac{gq M_P}{m} 
> \frac{gq M_P}{\Lambda} 
\gtrsim \frac{g q M_P}{ g M_P} = q \geq 1 \, .
\label{eq:light}
\end{equation}
The second inequality follows from 
the bound on the UV cut-off scale 
\eqref{eq:cutWGC},
while the last inequality follows from 
the convention used in this article\footnote{%
The 
physical
result of course does not depend on
the convention used.
One just need to use the same convention
which was used to state the magnetic WGC
\eqref{eq:cutWGC}.} 
that the smallest charge known to the 
low energy observers associated with 
the EFT under consideration is one.
Thus any particle whose mass is below 
the UV cut-off scale $\Lambda$
satisfies the electric WGC bound \eqref{eq:eWGC},
once $\Lambda$ satisfies the bound \eqref{eq:cutWGC}.
To put it differently:
{\em When one does not consider charged particles whose mass
is above the UV cut-off scale satisfies the bound
\eqref{eq:cutWGC},
the electric WGC is trivially satisfied.}
On the other hand, when there is no charged particles
below the UV cut-off scale of an EFT,
unless one considers charged particles whose mass is 
above the UV cut-off scale,
{\em \eqref{eq:cutWGC} itself cannot be stated},
since there is no physically meaningful way
to define the gauge coupling appearing in \eqref{eq:cutWGC}.
These observations give strong motivation to study
the case when some of the stable heavy charged particles
are known to the low energy observers associated with EFTs,
which is the main theme of the current article.

\subsection{Comments on monopoles}\label{subsec:monopoles}

From eq.~\eqref{eq:mmono} it follows that
monopoles are much heavier than the 
UV cut-off scale $\Lambda$
in the perturbative regime $g \ll 1$.
However,
there is a difference between
stable heavy electrically charged particles
which satisfy the electric WGC bound \eqref{eq:eWGC}
and monopoles.
If no particles whose
mass is below the UV cut-off scale $\Lambda$
of the EFT under consideration
satisfies the electric WGC bound
\eqref{eq:eWGC},
the electric WGC predicts
the existence of a charged particle
whose mass is above $\Lambda$ 
and which satisfies the electric WGC bound.
But other than the electric WGC itself,
there is no reason to expect the existence of such particle.
On the contrary,
monopoles are classical solutions
of Abelian gauge theories,
therefore their existence
is expected once the EFT 
with an Abelian gauge symmetry is given.
Since the low energy observers
are not guaranteed to know the true
smallest charge, there is also a limitation
in their knowledge 
of possible magnetic charges of the monopoles.
Recall that
I am using the convention that the smallest 
electric charge known to the low energy observers
is normalized to one.
In this convention, together with 
the normalization of the magnetic gauge coupling 
constant \eqref{eq:gm},
the low energy observers predict
the existence of magnetic monopoles with 
magnetic charge $q_m$ being non-zero integers:
\begin{equation}
q_m = \pm 1, \pm 2, \pm 3, \cdots \,.
\label{eq:m}
\end{equation}
However,
if the true smallest electric charge
is fractional, say $1/3$,
then the magnetic charges allowed by
the Dirac quantization condition
are integers multiple of three:
\begin{equation}
q_m = \pm 3, \pm 6, \pm 9, \cdots \, .
\label{eq:3m}
\end{equation}
Therefore,
what the low energy observers should expect 
is that the magnetic monopoles exist, 
and their magnetic charge
should be \textit{some} integer.
But they should not expect
that the magnetic monopoles with 
all integer charges must exist.
In particular, they should not assume that
the magnetic monopole with the unit charge must exist.
This is because such prediction can fail 
if there exists a particle with a smaller charge
than that known to the low energy observers,
as in the example above.
The limited knowledge of the low energy observers
leads to a looser bound on the UV cut-off scale of 
the EFT under consideration,
but a looser bound will not be wrong,
it is just looser than the bound
one can get from more precise knowledge.

If the low energy observers
detect a monopole and get to know its magnetic charge,
they can be at least sure for the existence of 
that magnetic charge.
They can also constrain possible electric charges
by the Dirac quantization condition.
In the rest of this article,
I focus on the cases
in which monopoles
are not observed by the low energy observers.
The cases in which
some monopoles have been detected
by the low energy observers
can be studied in a way
similar to the analysis below.

\subsection{BH discharge arguments 
and mass of charged particles}\label{subsec:BH}

The discharge process of BHs
provided an argument for the WGC \cite{ArkaniHamed:2006dz}.
I will examine it below.
The main conclusion I draw here
is that 
no restriction on the mass of 
the charged particles 
is put from this argument.\footnote{%
The same conclusion was also stated 
in \cite{Saraswat:2016eaz},
but here I give more detailed analysis 
of the actual discharge processes.}
In particular, the mass of the charged particle can be
above the UV cut-off scale of the EFT in which one examines the WGC.
This conclusion provides another reason for including
stable heavy particles whose mass is above
the UV cut-off scale of the EFT
into consideration when discussing the WGC.

The condition that
charged BHs can release more charges than mass
in Planck unit
was one of the physical motivations
behind the original proposal of the WGC \cite{ArkaniHamed:2006dz}.
In \cite{ArkaniHamed:2006dz}, the main concern was
whether charged BHs were kinematically allowed to decay.
The actual decay channels
in semi-classical regime 
have been
examined in \cite{Banks:2006mm} in the context of the WGC,
extending the work of \cite{Gibbons:1975kk}.
As the interest of the current article is
the low energy observer's perspective,
it would be adequate to assume that
the UV cut-off scale of EFTs considered
is at least few orders below the Planck scale,
thus the semi-classical analysis will be sufficient.

Let me start with a brief review of the discharge process of BHs
\cite{Gibbons:1975kk,Banks:2006mm}.
To fix my convention, I write down
the Einstein action coupled to an Abelian gauge field:
\begin{equation}
S = 
\frac{1}{16 \pi G}
\int d^4 x
\sqrt{-\det [g_{\mu\nu}]}
\left[
R
-\frac{1}{4} F_{\mu\nu} F^{\mu\nu}
\right]
\, .
\label{eq:action}
\end{equation}
Here, the cosmological constant is assumed to be negligible.
Then, 
assuming naturalness,\footnote{%
For a review of naturalness in EFTs,
see for example \cite{Giudice:2008bi}.}
the action
\eqref{eq:action} gives the leading terms 
in the low energy approximation.
Charged BHs are described by
the Reissner-Nordstr\"om (RN) BH solution:
\begin{align}
ds^2 = -&f(r) dt^2 + \frac{1}{f(r)} dr^2 + r^2 d\Omega_2^2 \, ,
\quad
A_0(r) = \frac{g Q}{r} \, ,
\nn \\
&f(r) = 1 - \frac{2 G M}{r} + \frac{G g^2 Q^2}{r^2} \, ,
\label{eq:RN}
\end{align}
where $Q$ and $M$ are the charge and the mass of the BH,
respectively.
I require that
the charge and the mass of BHs
satisfy the BPS bound:\footnote{%
This requirement is motivated
by the cosmic censorship hypothesis,
which may be related to the WGC
\cite{Crisford:2017gsb}.}
\begin{equation}
\frac{g Q M_P}{M} \lesssim 1 \, .
\label{eq:BPS}
\end{equation}
The radius of the outer horizon of
the black hole is given by
\begin{equation}
r_+ = G M + \sqrt{(GM)^2 - Gg^2Q^2} \, .
\label{eq:BHr}
\end{equation}
Notice that $GM \leq r_+ \leq 2 GM$,
so $r_+$ is of the order of $GM$.

Let $q$ and $m$ be the charge and the mass of
a particle which satisfy
the electric WGC bound \eqref{eq:eWGC}.
When the temperature of the BH is high
compared with the mass of the charged particle,
Hawking radiation is the dominant process for 
the discharge of the BH,
while when the temperature is lower than the mass,
the Schwinger pair-production is the dominant process.

\subsubsection*{Discharge by Hawking 
radiation}\label{subsubsec:Hawking}

I first consider the case where
the Hawking temperature $T_H$ of a BH
is higher than the mass $m$ of the 
a particle with charge $q$ satisfying
the electric WGC bound \eqref{eq:eWGC}:
\begin{equation}
T_{H} \gtrsim m \, .
\label{eq:hot}
\end{equation}
Since for a BH with given mass $M$
\begin{equation}
\frac{1}{GM} \gtrsim T_{H} \, ,
\label{eq:MT}
\end{equation}
I obtain
\begin{equation}
\frac{1}{GM} \gtrsim m \, ,
\label{eq:Mmi}
\end{equation}
which can be rewritten as
\begin{equation}
M \lesssim \frac{M_P^2}{m} \, .
\label{eq:MbH}
\end{equation}

The above was a condition for the BH to 
efficiently emit
particles with mass $m$.
In order for the charge to be emitted efficiently,
the chemical potential should be greater than the mass of the 
charged particle.
This condition gives
\begin{equation}
\frac{g^2 q Q}{r_+} \gtrsim m \, .
\label{eq:chem}
\end{equation}
Since $r_+ \sim GM \sim M/M_P^2$ and 
$gQM_P \leq M$ from \eqref{eq:BPS},
it follows that
\begin{equation}
\frac{g q M_P}{m} \gtrsim 1 \, .
\label{eq:qm}
\end{equation}
This is nothing but the electric WGC bound
\eqref{eq:eWGC}, which I already assumed
so that the charged particle can take
more charge than mass in Planck unit
from the BH.

\subsubsection*{Discharge by Schwinger 
pair-productions}\label{subsubsec:Schwinger}

Next I consider the case 
when the Hawking temperature of the 
BH is below the mass of
a particle that satisfies
the electric WGC bound \eqref{eq:eWGC}:
\begin{equation}
 T_H \lesssim m \, ,
\label{eq:cold}
\end{equation}
where $m$ is the mass of the particle.
The dominant process in this case is the
Schwinger 
pair-production.

A RN BH discharges appreciably
through the pair creations of charged particles 
with charge $q$ and mass $m$
by the Schwinger pair-productions when
\cite{Gibbons:1975kk,Banks:2006mm} 
\begin{equation}
m^2 \lesssim \frac{g^2 q Q}{r_+^2} \, .
\label{eq:mdc}
\end{equation}
Since $r_+ \sim GM \sim M/M_P^2$ and 
$gQM_P\lesssim M$ by \eqref{eq:BPS},
it follows that
\begin{equation}
m^2 \lesssim g q \frac{M_P^3}{M} \, .
\label{eq:m2}
\end{equation}
This can be rewritten as
\begin{equation}
M \lesssim \frac{M_P^2}{m} \left(\frac{gq M_P}{m} \right) \, .
\label{eq:MbSch}
\end{equation}

\subsubsection*{The bound on the mass of BHs
which can efficiently discharge}\label{subsubsec:Mbound}

From \eqref{eq:MbH} and \eqref{eq:MbSch},
it follows that in order for a BH to discharge 
efficiently,
there are bounds on the mass of the BH
\cite{Gibbons:1975kk}.
If one requires RN BHs with arbitrary large mass 
should decay efficiently, 
one need to have a charged particle
with arbitrary large ratio $M_P/m$.
This directly follows from \eqref{eq:MbH}
when the Hawking radiation is the dominant process.
For the case when the Schwinger pair-production
is the dominant process for discharge,
note that $g q$ cannot be infinite,
i.e. the notion of an infinite charge would be ill-defined,
and in fact
$g q \lesssim 1$ is required
for the gauge theory to be in the perturbative regime,
which I will assume in the following.
Then, from \eqref{eq:MbSch}
the condition that BHs with arbitrary mass can decay 
via the Schwinger process requires arbitrary large ratio $M_P/m$.
This is equivalent to requiring the existence of a
massless charged particle,
which appears to be a too strong constraint
which even QED
in our universe does not satisfy.
This would not be what the WGC
requires on EFTs.
Notice that the BHs with arbitrary mass should 
discharge efficiently 
led to a requirement for IR 
of the theory (i.e. the smallest mass of 
the charged particles which 
satisfy the electric WGC bound is relevant)
rather than UV.

If the WGC does not require
BHs with arbitrary mass to decay efficiently, 
there would be two alternative 
possibilities for 
the requirement for the BH discharge processes
in the argument for the WGC:
\begin{enumerate}
	\item BHs need not decay efficiently.
				Both Hawking radiation and 
				Schwinger pair-production can be
				interpreted as
				quantum tunneling 
				\cite{Parikh:1999mf,Brezin:1970xf},
				and as such
				as long as the final state is
				kinematically allowed,
				the BH has a probability to decay,
				though it is exponentially suppressed
				and the time scale required would easily exceed
				the age of the Universe.
  \item Only BHs smaller than a certain scale,
	      possibly a certain particle physics scale,
	      are required to decay efficiently.
\end{enumerate}
The first possibility does not impose any further condition
on the mass of the charged particle which satisfy
the electric WGC bound \eqref{eq:eWGC}.
I think this is indeed reasonable.

The second possibility 
requires an extra assumption
about what should be the ``certain scale" which bounds
the size of the BHs which should decay efficiently.
While some arguments to avoid
remnants\footnote{For a review of BH remnants and 
the pathologies they may introduce, 
see for example
\cite{Giddings:1995gd}.} 
may provide such an additional input,
I have not come up with a reasonable candidate for such a scale.
To get the feeling of the scales involved, 
consider
QED 
with the electron as the lightest charged particle,
which obviously satisfies 
the electric WGC bound \eqref{eq:eWGC}.
Then, the bound \eqref{eq:MbSch} gives
$M \lesssim 10^5 M_{\odot}$,
where $M_{\odot}$ is the solar mass
\cite{Gibbons:1975kk}.
However, this astrophysical mass
is due to the large
hierarchy between the Planck scale
and the mass of the electron.
In a model in which 
the charged particles relevant for 
the discharge of a BH have mass 
just few orders below the Planck scale,
the mass and the size of BHs
which can efficiently discharge 
can easily fall in the 
realm of particle physics.

In the rest of this article, I take the first possibility,
i.e. I assume that 
BHs only need to be kinematically allowed to discharge,
the actual process for the discharge need not be efficient.
Then, as stated above,
there is no constraint on the mass of the charged particles
which satisfy the electric WGC bound \eqref{eq:eWGC}.
An important consequence is that
{\em the charged particles relevant for BH discharge
need not have mass below the UV cut-off scale of an EFT
under consideration.}
The above observation supports the 
idea 
that one should take into account 
stable heavy particles
whose mass is above the UV cut-off scale of EFTs
when studying constraints of the WGC to 
low energy EFTs.

\section{The Saraswat model}\label{sec:srswt}

\subsection{The WGC under a 
spontaneous gauge symmetry breaking}\label{subsec:srswt}

In an interesting article 
\cite{Saraswat:2016eaz},
Saraswat proposed a simple model in which
the WGC is respected at high energy, but
after a spontaneous gauge symmetry breaking
the strong version of
the electric WGC can be violated 
at low energy.\footnote{%
The low energy theory is similar to the one studied in
\cite{Hebecker:2015rya}.}
The embedding of the Saraswat model to string theory 
was examined in \cite{Ibanez:2017vfl}.
Below I will further examine the cases
where the weak version of the 
electric WGC is not actually violated,
but \textit{appears} to be violated
for the low energy observers
associated with the low energy EFT,
due to the limitation of their knowledge
regarding the stable heavy particles.
I will also show that the magnetic WGC 
is not violated at low energy
in this model.

The Saraswat model has two Abelian gauge groups
which I denote $U(1)_A$ and $U(1)_B$.
Following \cite{Saraswat:2016eaz},
for simplicity I take 
$g_A = g_B = g$, 
where $g_A$ and $g_B$ are the coupling constants of
the gauge group $U(1)_A$ and $U(1)_B$, 
respectively.
It is straightforward to generalize the arguments below
to the case $g_A \ne g_B$.
As explained in the previous section,
these gauge couplings are
normalized so that 
the smallest charge known
to the low energy observers
associated with the EFT is one.
With this convention, I introduce 
fields $\psi_A$ 
and $\psi_B$ 
which have the charge vectors
$\vec{q}_A=(1,0)$ and $\vec{q}_B=(0,1)$, respectively.
Here, the first entry refers to the charge 
with respect to the $U(1)_A$ gauge group,
and the second entry refers to the charge with respect to
the $U(1)_B$ gauge group.
In addition, I introduce a Higgs field $H$ which has
a charge vector $\vec{q}_H=(Z,1)$,
where $Z$ is an integer. 
I assume that
all the particles appearing in this EFT
have mass below the UV cut-off scale $\Lambda$ of the EFT: 
\begin{equation}
m_A, m_B, m_H < \Lambda \, ,
\label{eq:mAmB}
\end{equation}
where $m_A$ and $m_B$ are the masses of the 
$\psi_A$ and $\psi_B$, respectively,
and $m_H$ is the mass of the Higgs particle
in the vacuum (to be explained below).
I further assume that
$\psi_A$ and $\psi_B$ are the lightest particles
among those charged with respect to
$U(1)_A$ and $U(1)_B$, respectively.
Since the Higgs field is charged under both $U(1)_A$ and $U(1)_B$,
this assumption implies
\begin{equation}
m_A, m_B < m_H \, .
\label{eq:lightest}
\end{equation}
I require that the model satisfies the 
strong electric- and the magnetic WGC
to begin with, and examine 
what happens below
the spontaneous gauge symmetry breaking scale.
When there are multiple $U(1)$ gauge groups,
the bounds of the WGC need to be generalized
accordingly
\cite{Cheung:2014vva}.
To describe the WGC bounds,
I define the charge-to-mass ratio vectors as follows:
\begin{equation}
\vec{z}_A =
\frac{g\vec{q}_AM_P}{m_A}
\, ,\quad
\vec{z}_B =
\frac{g\vec{q}_BM_P}{m_B} \, .
\label{eq:zAzB}
\end{equation}
The strong electric WGC 
means the the electric WGC 
is satisfied by the lightest charged particles
$\psi_A$ and $\psi_B$.
Then, 
the electric WGC
requires that
a unit disk, which represents
possible charge-to-mass ratio of BHs
constrained by the BPS bound \eqref{eq:BPS},
is contained in the convex hull
spanned by the vectors $\pm \vec{z}_A$ and $\pm \vec{z}_B$
\cite{Cheung:2014vva}
(see also Appendix~\ref{notviolated}).
This requirement gives the conditions\footnote{%
In this article
I have been neglecting $\Ord(1)$ numerical factors,
therefore $1/\sqrt{2}$ 
in \eqref{eq:emAmB} and \eqref{eq:LambdaUV}
should be understood
only relative to \eqref{eq:eWGC}.}
\begin{equation}
m_A, m_B \lesssim 
\frac{g M_P}{\sqrt{2}} \, .
\label{eq:emAmB}
\end{equation}
The bound on the UV cut-off scale of the EFT
becomes
\begin{equation}
\Lambda\lesssim \frac{g M_P}{\sqrt{2}}  \, ,
\label{eq:LambdaUV}
\end{equation}
where $\Lambda$ is the UV cut-off scale
above which the description by this EFT breaks down.
Note that the assumptions 
made above are not all independent:
From 
the assumption \eqref{eq:mAmB}
and
the bound on the UV cut-off
\eqref{eq:LambdaUV},
the electric WGC \eqref{eq:emAmB} automatically follows,
as explained in Sec.~\ref{subsec:trivial}.

After the Higgs field acquires a
vacuum expectation value,
only a linear combination
of the gauge fields remains massless.
I call
the gauge field of the gauge group $U(1)_A$
as $A_\mu$,
and
the gauge field of the gauge group $U(1)_B$
as $B_\mu$.
Then, a convenient basis for identifying 
the massless combination of the gauge fields
is obtained by the linear transformation:
\begin{equation}
\left(
\begin{array}{c}
\tilde{A}_\mu\\
\tilde{B}_\mu
\end{array}
\right)
=
\left(
\begin{array}{cc}
\cos \gamma & \sin \gamma \\
-\sin \gamma & \cos \gamma
\end{array}
\right)
\left(
\begin{array}{c}
{A}_\mu\\
{B}_\mu
\end{array}
\right) \, ,
\label{eq:rot}
\end{equation}
where
\begin{equation}
\cos \gamma := \frac{Z}{\sqrt{1+Z^2}} \, ,
\quad
\sin \gamma := \frac{1}{\sqrt{1+Z^2}}\, .
\label{eq:gamma}
\end{equation}
Let $v$ be the vacuum expectation value of the Higgs field.
Then, $\tilde{A}_\mu$ acquires a mass 
$m_V = g \sqrt{1+Z^2}\, v$
while
$\tilde{B}_\mu$ remains massless.
I will refer to the unbroken $U(1)$ gauge symmetry group 
as $U(1)_{\mathrm{eff}}$,
and the broken one as $U(1)_{\mathrm{broken}}$.
A field which had a charge vector 
$(m,n)$ in the original basis
has 
$(m \cos \gamma  + n \sin \gamma , 
- m \sin \gamma  + n \cos \gamma )_{\mathrm{new}}$ 
in the new basis after the linear transformation
\eqref{eq:rot}.
For example, the field $\psi_A$
which had $(1,0)$ in the original basis
has $(\cos \gamma,-\sin \gamma)_{\mathrm{new}}$ 
in the new basis
(before rescaling the gauge coupling constant
following my convention, which will be done below).
When $Z \gg 1$, $\sin \gamma \ll 1$,
i.e. $\psi_A$ has a very small charge
(before rescaling the gauge coupling constant)
with respect to $U(1)_{\mathrm{eff}}$.
Such small charge,
which will be translated 
into small gauge coupling constant
after the rescaling following 
the convention
for the gauge coupling constant,
is a potential source
for a violation of the electric WGC.
I will study the consequence 
of this small charge in some detail below.

Now, I examine
the WGC from the viewpoint of 
the low energy observers
associated with the EFT with unbroken
$U(1)_{\mathrm{eff}}$ gauge symmetry.
I first need to identify what is the appropriate EFT
to state the WGC with respect to 
the $U(1)_{\mathrm{eff}}$ gauge symmetry.
For the clarity of the argument,
I call the EFT I started with as EFT$_{\mathrm{UV}}$,
and the EFT for the $U(1)_{\mathrm{eff}}$ gauge symmetry
as EFT$_{\mathrm{eff}}$.
Then, the UV cut-off scale $\Lambda_{\mathrm{eff}}$
for the EFT$_{\mathrm{eff}}$
should be at least below the 
gauge symmetry breaking scale
which is characterized 
by the mass of the
massive gauge field,
since otherwise EFT$_{\mathrm{UV}}$
with 
the $U(1)_A \times U(1)_B$ gauge symmetry
becomes the appropriate description.
Hence
\begin{equation}
\Lambda_{\mathrm{eff}} < m_V = g \sqrt{1 + Z^2} \, v  \, .
\label{eq:NPS1}
\end{equation}
However, I will obtain a tighter 
bound on $\Lambda_{\mathrm{eff}}$ 
from the analysis of the size of 
the monopole of $U(1)_{\mathrm{eff}}$
in this model below.

It is important to
notice that the WGC for EFT$_{\mathrm{UV}}$
and that for EFT$_{\mathrm{eff}}$ are different,
for example, the gauge groups and the gauge coupling constants
are different.

As I explained in Sec.~\ref{subsec:trivial},
when the mass of the field $\psi_A$ (or $\psi_B$)
is below the UV cut-off scale of EFT$_{\mathrm{eff}}$
i.e. 
$m_A < \Lambda_{\mathrm{eff}}$ (or $m_B < \Lambda_{\mathrm{eff}}$),
the electric WGC bound is
automatically satisfied by $\psi_A$ (or $\psi_B$)
in EFT$_{\mathrm{eff}}$.
The magnetic WGC in these cases can be analyzed 
in the same way as in the case 1 below.
In the following
I examine the remaining case $m_A, m_B > \Lambda_{\mathrm{eff}}$.
The Higgs particles can be stable
when $m_H < Z m_A + m_B$,
which could mostly be the case when $Z \gg 1$.
For simplicity,
I restrict myself to the cases
where the low energy observers associated with 
EFT$_{\mathrm{eff}}$
do now know the existence of the Higgs particles.
It is straightforward to include the case
where the existence of the Higgs particles
is known to the low energy observers
associated with EFT$_{\mathrm{eff}}$.
In the following, 
I will analyze the cases 
differ in 
which particles are known to the 
low energy observers associated
with EFT$_{\mathrm{eff}}$.

\subsubsection*{Case 1: Both $\psi_A$ and 
$\psi_B$ are known to the low energy observers
associated with EFT$_{\mathrm{eff}}$}

In this case, 
the $\psi_A$ particles
have the smallest charge 
with respect to
$U(1)_{\mathrm{eff}}$
among the charged particles
which are known to the low energy observers
associated with the EFT$_{\mathrm{eff}}$.
Thus in my convention that
the smallest charge known to the low energy observer is one,
the gauge coupling constant 
$g_{\mathrm{eff}}$
for the $U(1)_{\mathrm{eff}}$
gauge group is given as
\begin{equation}
g_{\mathrm{eff}} = g \sin  \gamma \, .
\label{eq:geff}
\end{equation}

\begin{figure}[htbp]
\centering
\includegraphics[width=5in]{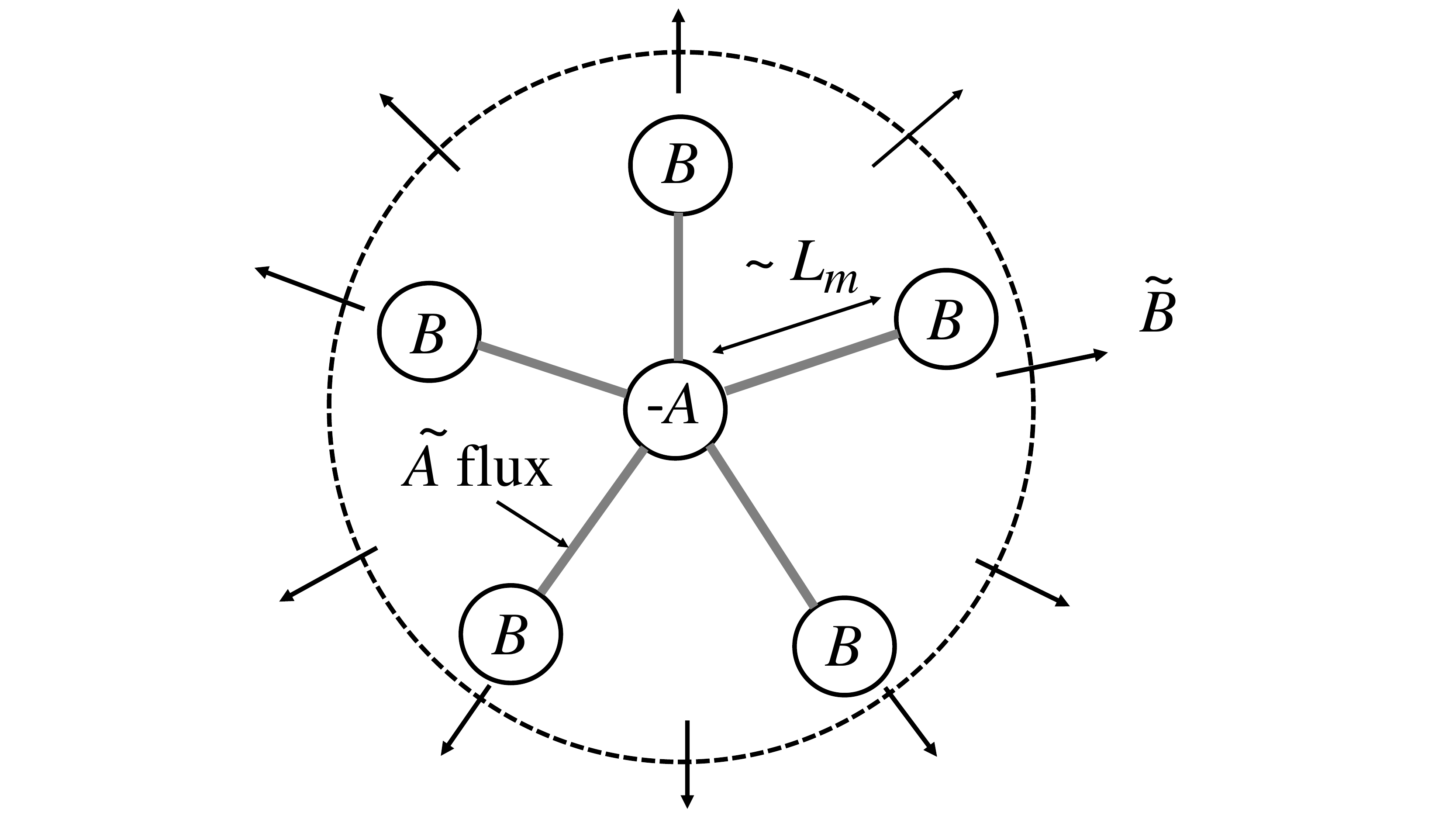} 
\caption{A schematic picture of a monopole
of $U(1)_{\mathrm{eff}}$ with unit magnetic charge.
It consists of 
one anti-monopole of $U(1)_A$ and
$Z$ monopoles of $U(1)_B$,
connected by the Nielsen-Olesen flux tubes
with $1/(1+Z^2)$ of the 
$U(1)_{\mathrm{broken}}$ flux
which I indicate by $\tilde{A}$
inside of each \cite{Saraswat:2016eaz}.
Here, the flux number is normalized
so that that from
a monopole with unit magnetic charge
is normalized to one.
\label{fig:monopole}}
\end{figure}

\begin{figure}[htbp]
\centering
\includegraphics[width=4in]{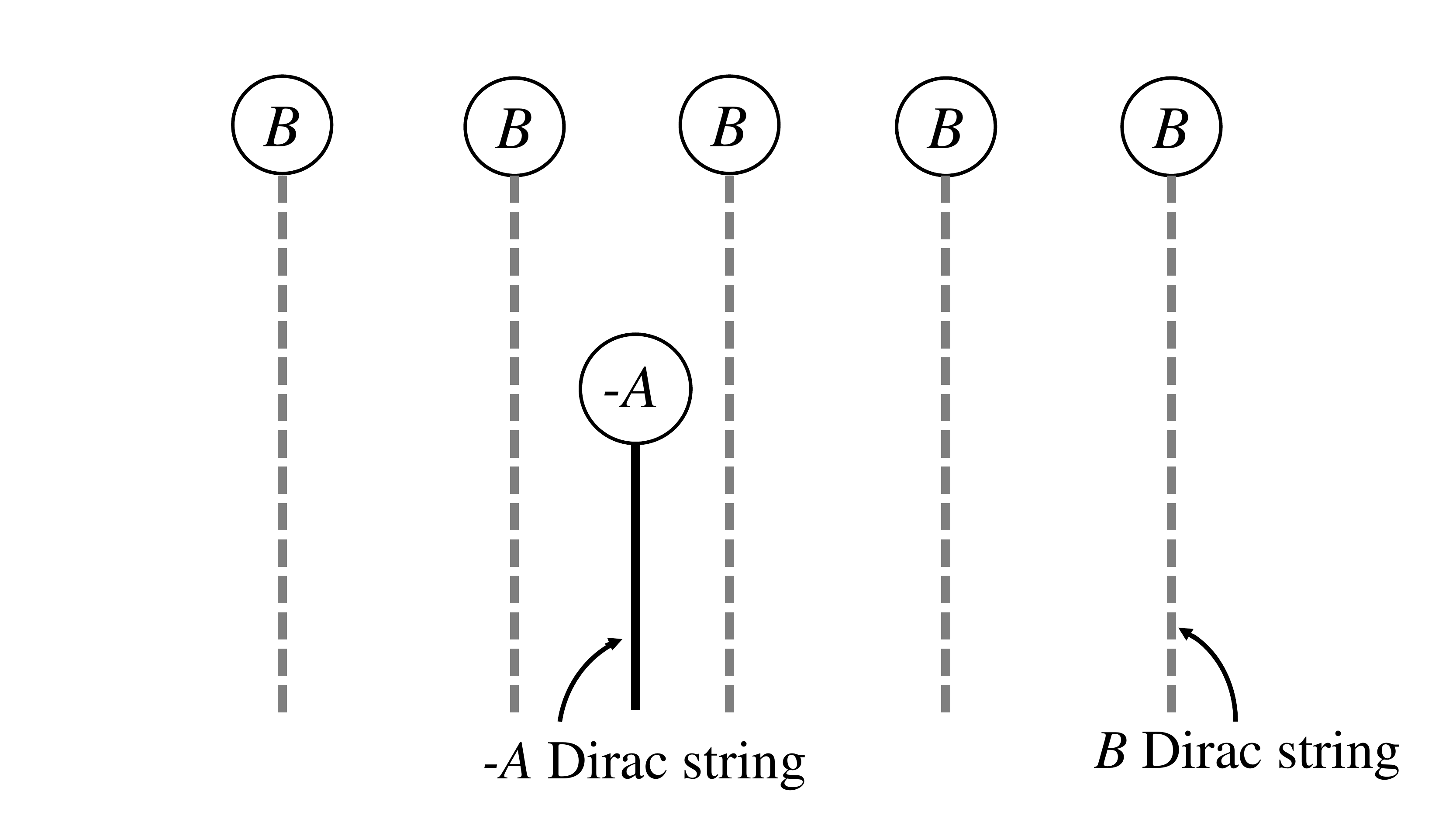} 
\caption{An anti-monopole of $U(1)_A$ and 
$Z$ monopoles of $U(1)_B$
in the unbroken phase of the
$U(1)_A \times U(1)_B$ gauge theory.
\label{fig:DiracString}}
\end{figure}

\begin{figure}[htbp]
\centering
\includegraphics[width=4in]{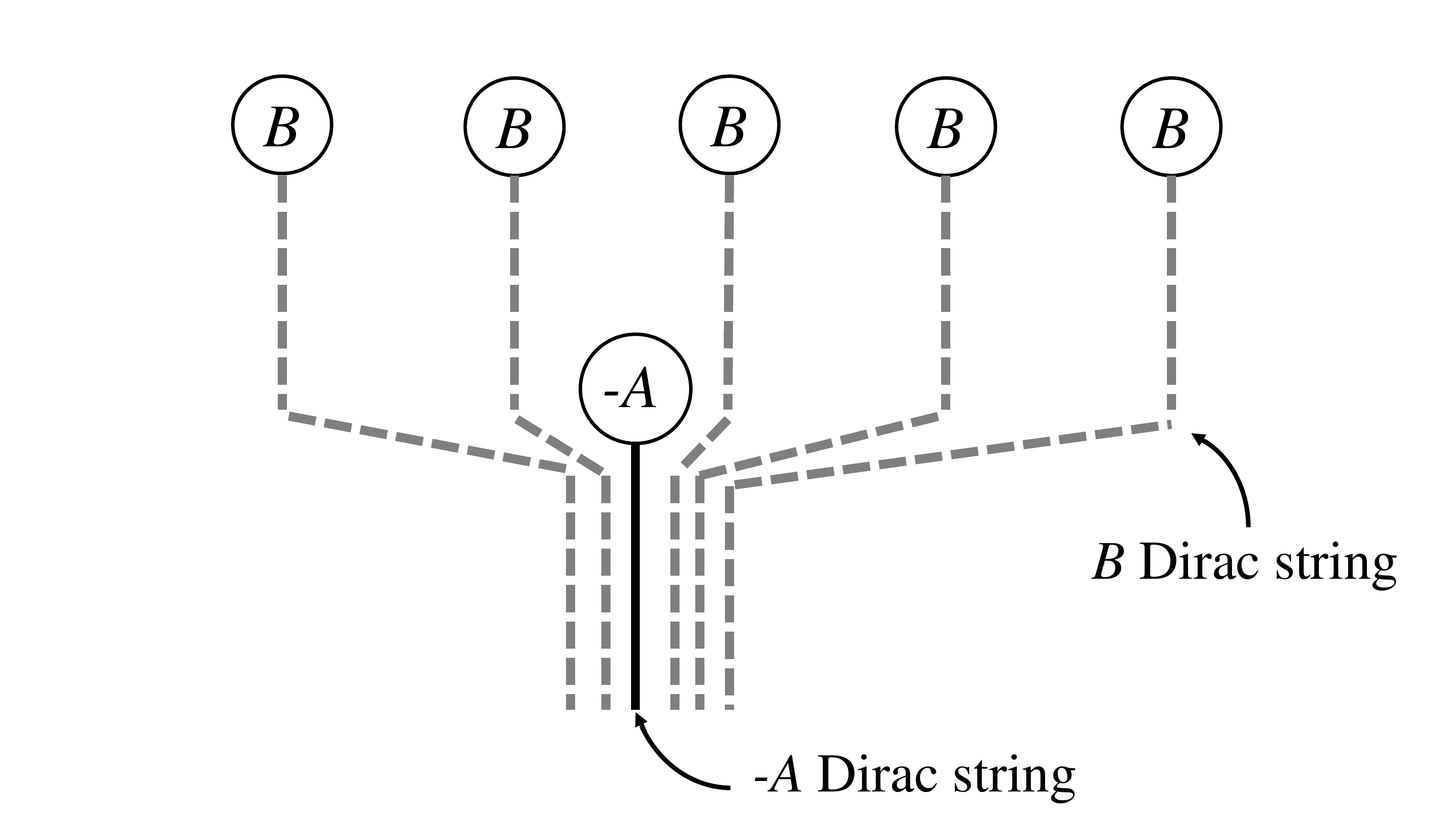} 
\caption{The monopoles in the broken phase of the gauge theory,
but Dirac strings still in the basis of $U(1)_A \times U(1)_B$.
While Dirac strings are not physical,
the magnetic flux of the broken gauge group 
$U(1)_{\mathrm{broken}}$ is physical.
One can choose a gauge in which
the Dirac string of $U(1)_{\mathrm{broken}}$
is along the magnetic flux
\cite{Creutz:1974vk,Ripka:2003vv}.
\label{fig:DiracString2}}
\end{figure}

\begin{figure}[htbp]
\centering
\includegraphics[width=4in]{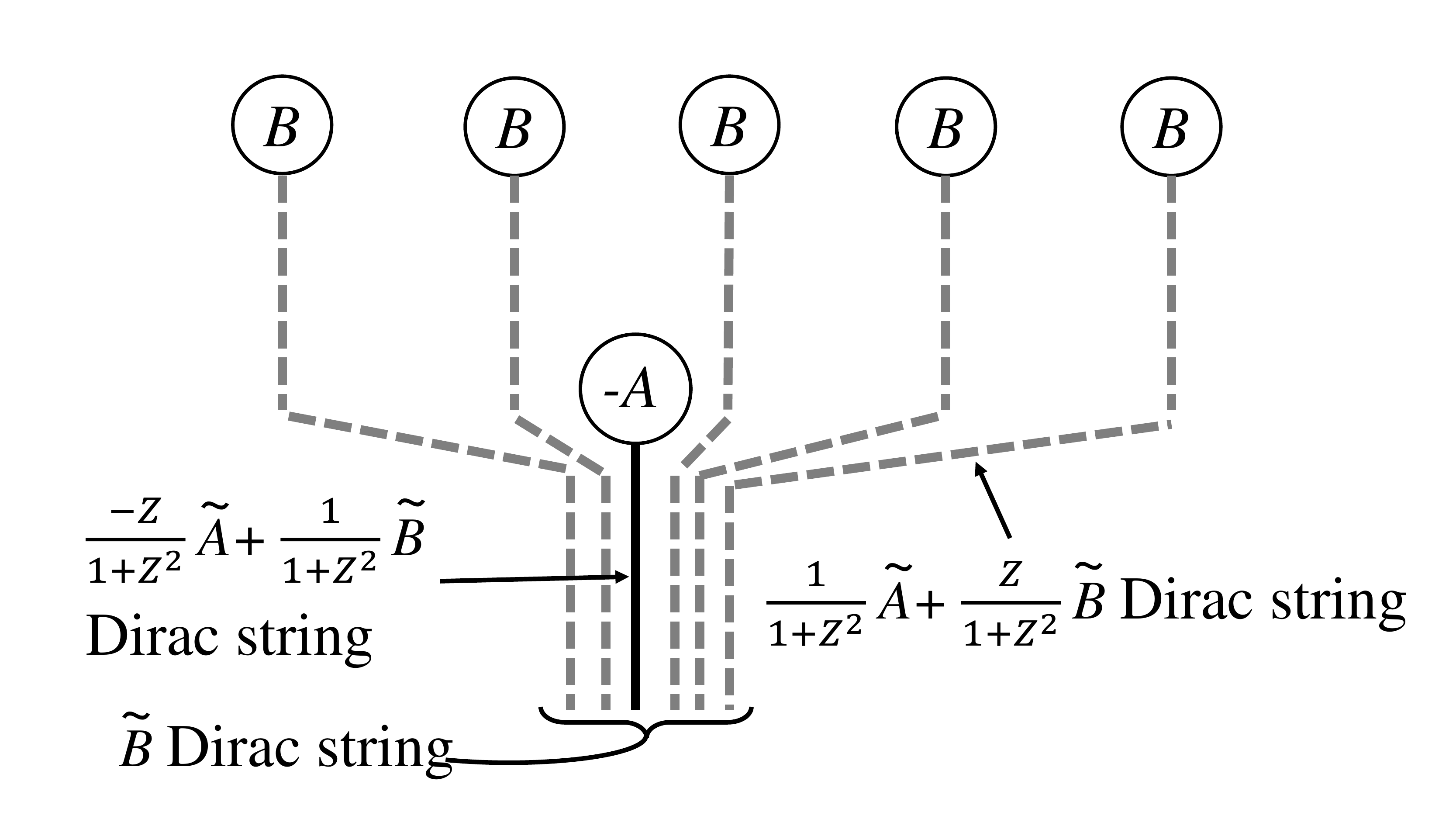} 
\caption{The same system as in Fig.~\ref{fig:DiracString2},
but in terms of the new basis 
$U(1)_{\mathrm{broken}} \times U(1)_{\mathrm{eff}}$.
The Dirac string of $U(1)_{\mathrm{broken}}$ is indicated
by $\tilde{A}$ and
the Dirac string of $U(1)_{\mathrm{eff}}$
is indicated by $\tilde{B}$.
The unit of the Dirac string
of $U(1)_{\mathrm{broken}}$ 
($U(1)_{\mathrm{eff}}$) 
is normalized so that 
that from the monopole 
with unit magnetic charge 
with respect to $U(1)_{\mathrm{broken}}$ 
($U(1)_{\mathrm{eff}}$) is one.
\label{fig:DiracStringNew}}
\end{figure}

\begin{figure}[htbp]
\centering
\includegraphics[width=4in]{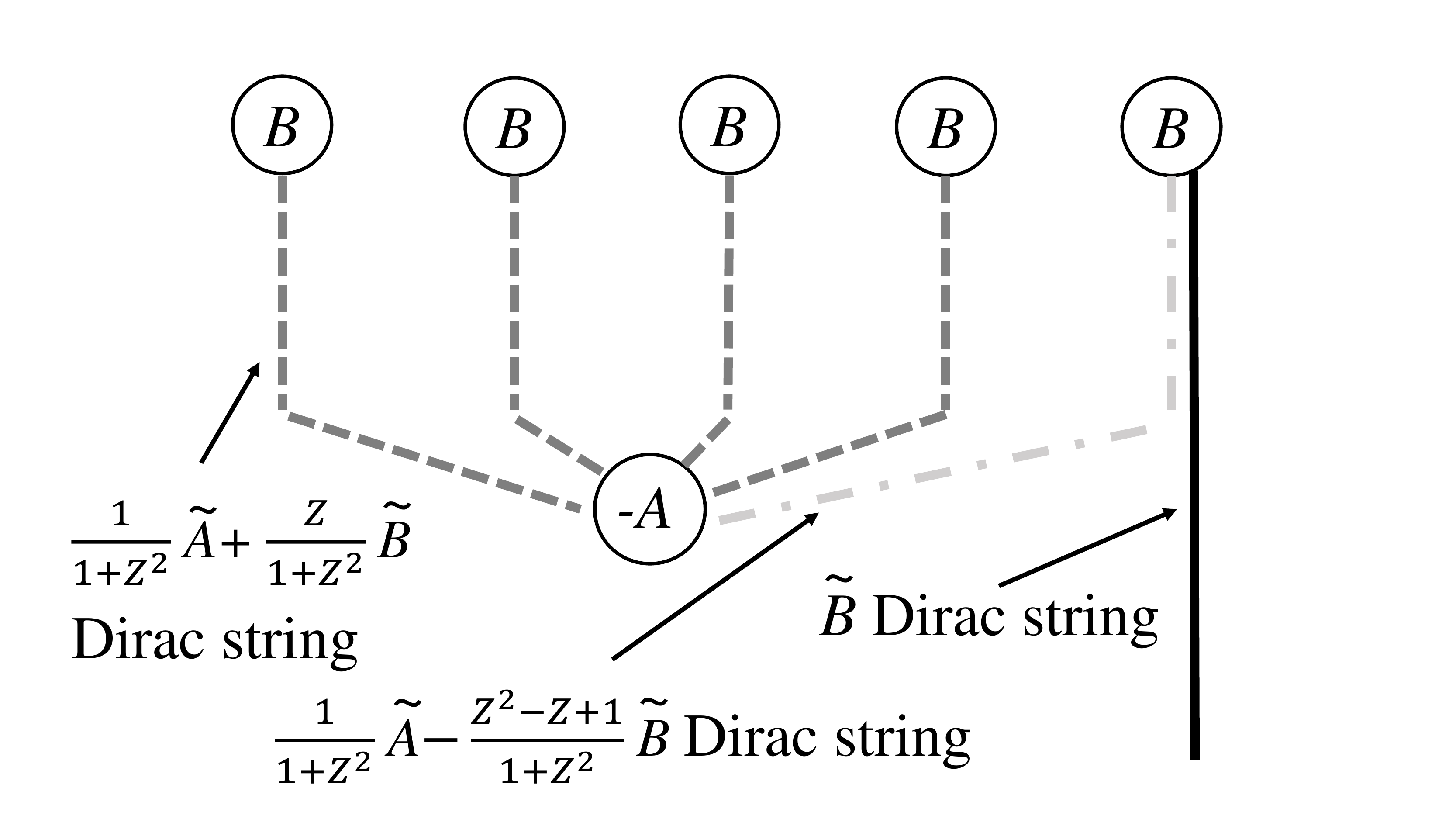} 
\caption{The same system as in Fig.~\ref{fig:DiracStringNew},
but in a different gauge.
A gauge transformation in $U(1)_{\mathrm{eff}}$
can move around one unit of the Dirac string of $U(1)_{\mathrm{eff}}$.
\label{fig:DiracStringNew2}}
\end{figure}

As mentioned earlier,
it is crucial to 
identify the 
UV cut-off scale $\Lambda_{\mathrm{eff}}$
of EFT$_{\mathrm{eff}}$ correctly
in order to discuss the magnetic WGC.
As it turns out, for this purpose
it is important to first understand
how the monopole of $U(1)_{\mathrm{eff}}$
is described in the UV theory,
i.e. EFT$_{\mathrm{UV}}$.
As found in \cite{Saraswat:2016eaz},
a monopole with unit magnetic charge
with respect to $U(1)_{\mathrm{eff}}$
is constructed 
in EFT$_{\mathrm{UV}}$
from
$Z$ monopoles of $U(1)_B$
and one anti-monopole of $U(1)_A$,
connected by Nielsen-Olesen flux tubes
\cite{Nielsen:1973cs,Nambu:1974zg}
(see Fig.~\ref{fig:monopole}).
Indeed, with the magnetic gauge coupling
\eqref{eq:gm},
the magnetic charge vector for this system 
in the new basis 
can be read off using 
the linear transformation \eqref{eq:rot} as
\begin{equation}
g_m
\left(
\begin{array}{cc}
	\cos \gamma & \sin \gamma \\
	-\sin \gamma & \cos \gamma
\end{array}
\right)
\left(
\begin{array}{c}
	-1 \\
	 Z
\end{array}
\right)
=
\frac{2\pi}{g}
\sqrt{1+Z^2}
\left(
\begin{array}{c}
	0 \\
	1
\end{array}
\right)_{\mathrm{new}}
=
\frac{2\pi}{g_{\mathrm{eff}}}
\left(
\begin{array}{c}
	0 \\
	1
\end{array}
\right)_{\mathrm{new}}\, .
\label{eq:magc}
\end{equation}
The radius of the flux tube is
$\sim 1/m_V$,
and the tension is $\sim v^2$.
Here, as in \cite{Saraswat:2016eaz},
I assumed that the mass of the Higgs field is greater
than the mass $m_V = g \sqrt{1+Z^2}\, v$ of the massive gauge field,
in which case the symmetry-breaking vacuum 
acts like a type II superconductor
\cite{Nielsen:1973cs,Nambu:1974zg}.
This assumption is consistent 
with the assumptions I made and will make later,
$v$, $m_H < \Lambda$ and $g Z \lesssim 1$.
The size $L_m$ of the system
is estimated
from a balance between
the energy in the flux tubes
$\sim Z v^2 L_m$
and the magnetic repulsive potential energy between the monopoles
$\sim Z^2 \left(\frac{2\pi}{g}\right)^2 \frac{1}{L_m}$:
\begin{equation}
L_m \sim \frac{\sqrt{Z}}{g v} \, .
\label{eq:Lm}
\end{equation}
Here, I did not take into account 
the gravitational force, 
as it will turn out to be
sub-dominant
due to the magnetic WGC being satisfied.
The mass $m_{m0}$ of the 
monopole with
unit magnetic charge with respect to the gauge group
$U(1)_{\mathrm{eff}}$
is estimated from the energy in the flux-tubes
and the magnetic potential:
\begin{equation}
m_{m0}  \sim \frac{Z^{\frac{3}{2}} v}{g}
\, .
\label{eq:Mmonoeff}
\end{equation}
Since the internal structure of 
the monopole of $U(1)_{\mathrm{eff}}$ 
cannot be described by
EFT$_{\mathrm{eff}}$,
the size of the monopole $L_m$
represents the scale where
the description by EFT$_{\mathrm{eff}}$ breaks down.
In fact, 
consider the low energy observers
who have access to energy scale beyond
$1/L_m$ but below $m_V$.
Since they
can probe beyond the length scale shorter than $L_m$,
they can observe
the monopoles of $U(1)_B$ and
anti-monopole of $U(1)_A$
individually 
using the gauge field of $U(1)_{\mathrm{eff}}$.
Each monopole of $U(1)_B$ has magnetic charge
$Z/(1+Z^2)$ with respect to $U(1)_{\mathrm{eff}}$,
while 
the anti-monopole of $U(1)_A$ has $1/(1+Z^2)$
(see Table~\ref{table:charges},
which can be calculated as in \eqref{eq:magc}).
\begin{table}
\begin{center}
\begin{tabular}{  | l | l | l | }
\hline
particle	& charge in $U(1)_A \times U(1)_B$ & 
charge in $U(1)_{\mathrm{broken}}\times U(1)_{\mathrm{eff}}$ \\
\hline
$\psi_A$ & (e) $g (1,0)$ & (e) $g_{\mathrm{eff}} (Z,-1)_{\mathrm{new}}$ \\
\hline
$\psi_B$ & (e) $g (0,1)$ & (e) $g_{\mathrm{eff}} (1,Z)_{\mathrm{new}}$ \\
\hline
Higgs $H$ & (e) $g (Z,1)$ & (e) $g_{\mathrm{eff}} (1+Z^2,0)_{\mathrm{new}}$ \\
\hline
$U(1)_A$ anti-monopole & (m) $\frac{2\pi}{g} (-1,0)$ & 
(m) $\frac{2\pi}{g_{\mathrm{eff}}} 
\left(-\frac{Z}{1+Z^2},\frac{1}{1+Z^2}\right)_{\mathrm{new}}$ \\
\hline
$U(1)_B$ monopole & (m) $\frac{2\pi}{g} (0,1)$ & 
(m) $\frac{2\pi}{g_{\mathrm{eff}}} 
\left(\frac{1}{1+Z^2}, \frac{Z}{1+Z^2} \right)_{\mathrm{new}}$\\
\hline
\end{tabular}
\caption{(e) and (m) indicate the electric- and magnetic charge,
respectively. 
I also associated the gauge couplings to the charges.
The subscript ``new'' indicates that they are in
the new basis of the charge vector
(the charge vector in $U(1)_{\mathrm{broken}}\times U(1)_{\mathrm{eff}}$ basis).
\label{table:charges}}
\end{center}
\end{table}
They can also observe the attractive
force between the monopoles of 
$U(1)_A$ and
$U(1)_B$
due to the tension of the flux tubes,
although they cannot resolve
the radial size of the flux tube $\sim 1/m_V$.
The monopoles with
fractional magnetic charges with respect
to $U(1)_{\mathrm{eff}}$
must have observable Dirac strings
unless some new physics need to appear and
prevent the Dirac strings from being observed:
Stable heavy particles whose mass
is above the UV cut-off of the EFT
can be described in the framework of EFT
as reviewed in \ref{subsec:HPET},
and once $\Psi_A$ and $\Psi_B$ are described in this way,
Dirac string becomes observable through the Aharonov-Bohm Effect,
unless new physics is introduced above $L_m$.
Indeed, in the current case,
I started from the UV theory EFT$_{\mathrm{UV}}$,
and in EFT$_{\mathrm{UV}}$,
which is a new physics for EFT$_{\mathrm{eff}}$,
the fractional magnetic charges of the monopoles
are
consistent with the Dirac quantization,
as I explain below.
To make Dirac quantization with respect to 
$U(1)_{\mathrm{eff}}$ consistent,
the Dirac strings of 
$U(1)_{\mathrm{eff}}$ and $U(1)_{\mathrm{broken}}$
should be combined appropriately.\footnote{An 
analogous argument has been
given in order to make Dirac quantization condition
consistent with the fractional electric charge of quarks
with a Dirac monopole with unit magnetic charge.
For further readings, see
for example Sec.~5.1 of Ref.~\cite{Weinberg:1992hc}.}
Actually, the appropriate combination
is the Dirac strings of the original gauge group,
$U(1)_A \times U(1)_B$,
see Fig.~\ref{fig:DiracString}-\ref{fig:DiracStringNew}.
The Dirac string of 
$U(1)_{\mathrm{eff}}$
for the monopole of $U(1)_B$
corresponds to that for the
fractional magnetic charge $Z/(1+Z^2)$
(Table~\ref{table:charges}).
This should be combined 
with the Dirac string of $U(1)_{\mathrm{broken}}$
which corresponds to the 
fractional magnetic charge $1/(1+Z^2)$.
When a $\psi_A$ particle 
whose charge in 
$U(1)_{\mathrm{broken}}\times U(1)_{\mathrm{eff}}$ basis
is $(Z,-1)_{\mathrm{new}}$
travels around the combined Dirac string,
the Aharonov-Bohm phase from the coupling to $U(1)_{\mathrm{eff}}$
gauge field
cancels with that from the coupling to $U(1)_{\mathrm{broken}}$ gauge field:
\begin{equation}
g_{\mathrm{eff}} Z \cdot \frac{2\pi}{g_{\mathrm{eff}}} \frac{1}{1+Z^2}
+
g_{\mathrm{eff}} (-1) \cdot \frac{2\pi}{g_{\mathrm{eff}}} \frac{Z}{1+Z^2}
= 0 \, .
\label{eq:ABphase}
\end{equation}
Hence, the $\psi_A$ particle cannot probe the 
combined Dirac string.
Similarly,
the Dirac string from the anti-monopole of $U(1)_A$
should be a combination of that from
$-Z/(1+Z^2)$ of magnetic charge of $U(1)_{\mathrm{broken}}$ 
and
$1/(1+Z^2)$ of magnetic charge of $U(1)_{\mathrm{eff}}$.
One can check that neither
$\psi_A$ nor $\psi_B$
can probe the combined Dirac string
from the monopole of $U(1)_A$
or that from the monopole of $U(1)_B$.
One can choose a different gauge for
the unbroken gauge group $U(1)_{\mathrm{eff}}$.
However, 
only the Dirac string which corresponds to that for 
the unit magnetic charge monopole can be deformed
by the gauge transformation (Fig.~\ref{fig:DiracStringNew2}).
Such gauge transformation
only changes the Aharonov-Bohm phase
of the $\psi_A$ particle traveling 
around the Dirac string by $2\pi$, 
which is not observable.
Of course,
when I considered the monopoles
of $U(1)_A$ and $U(1)_B$,
I already took into account 
the Dirac quantization condition with respect to
these gauge groups,
therefore the Aharonov-Bohm phases of 
the charged particles 
going around the Dirac strings must come right.
Here, I gave a description in terms of 
the $U(1)_{\mathrm{broken}} \times U(1)_{\mathrm{eff}}$ basis.

Note that the physical 
magnetic flux in the Nielsen-Olesen flux tube
has the unit flux.
Since the Nielsen-Olesen flux tube
is a solution of the gauge-Higgs system,
the charge of the Higgs field determines 
the unit magnetic flux.
The Higgs field
has the charge with respect to the $U(1)_{\mathrm{broken}}$
(combined with the gauge coupling constant)
$g_{\mathrm{eff}}(1+Z^2)$ (see Table~\ref{table:charges}).
This leads to the quantization of the magnetic flux
in the Nielsen-Olesen flux tube as
\begin{equation}
\frac{2\pi n}{g_{\mathrm{eff}}(1+Z^2)}  \, \quad
\mbox{$n$: integer} \, .
\label{eq:fluxq}
\end{equation}
Here, the integer $n$ is the vorticity.\footnote{%
For a review of
Nielsen-Olesen flux tube (vortex)
and its topological property, 
see for example \cite{Weinberg:1992hc}.}

The Dirac string of $U(1)_{\mathrm{broken}}$
should not be confused with 
the magnetic flux of $U(1)_{\mathrm{broken}}$
inside the flux tube
\cite{Creutz:1974vk}.
The magnetic flux is physical
and there is no problem in
particles probing it through
the Aharonov-Bohm effect.
One can choose a gauge in which
the Dirac string of $U(1)_{\mathrm{broken}}$
can be placed along the magnetic flux by a choice of gauge
\cite{Ripka:2003vv}.

The existence of 
the gauge field of
$U(1)_{\mathrm{broken}}$ and the charges with respect to it
was crucial for making
the Dirac strings from
the monopoles of $U(1)_A$ and $U(1)_B$
with the fractional magnetic charges
with respect to the $U(1)_{\mathrm{eff}}$ gauge group
unobservable.
The gauge field of
$U(1)_{\mathrm{broken}}$ 
is a new physics beyond EFT$_{\mathrm{eff}}$.
While the massive gauge bosons of 
$U(1)_{\mathrm{broken}}$ 
do not appear as particle states
at the energy scale below $m_V$,
their existence is required
so that the Dirac strings not to be probed
through the Aharonov-Bohm effect.
Therefore, the size $L_m$ of the monopole 
of $U(1)_{\mathrm{eff}}$
is the scale where the description by 
EFT$_\mathrm{eff}$ breaks down.
Thus I identify 
the size of the monopole of $U(1)_{\mathrm{eff}}$
with 
the UV cut-off scale of 
EFT$_{\mathrm{eff}}$:\footnote{%
In the original article \cite{Saraswat:2016eaz},
the WGC with respect to the $U(1)_A \times U(1)_B$ gauge theory
and the WGC with respect to the $U(1)_{\mathrm{eff}}$ gauge theory 
were not clearly distinguished.
At $\Lambda_{\mathrm{eff}}$ which is the UV cut-off scale
for the $U(1)_{\mathrm{eff}}$ gauge theory,
the fractional magnetic charge with respect to 
the $U(1)_{\mathrm{eff}}$ gauge group makes 
the $U(1)_{\mathrm{eff}}$ gauge theory becomes theoretically inconsistent,
and the new degrees of freedom associated with 
the $U(1)_A \times U(1)_B$ gauge theory 
manifest themselves through the Aharonov-Bohm effect.}
\begin{equation}
\Lambda_{\mathrm{eff}} \sim \frac{1}{L_m} \sim \frac{g v}{\sqrt{Z}} \, .
\label{eq:Lambdaeff}
\end{equation}
Notice that this UV cut-off scale $\Lambda_{\mathrm{eff}}$
is much smaller than
the mass $m_V = g \sqrt{1+Z^2} \, v$ \eqref{eq:NPS1}
of the massive gauge field
when $Z \gg 1$.

With the identification of the UV cut-off scale 
in \eqref{eq:Lambdaeff},
the mass of the monopole of $U(1)_{\mathrm{eff}}$
\eqref{eq:Mmonoeff}
coincides with 
the general expectation for the 
mass of the Dirac monopole in EFT given in \eqref{eq:mmono}:
\begin{equation}
m_{m0} \sim \frac{\Lambda_{\mathrm{eff}}}{g_{\mathrm{eff}}^2} \, .
\label{eq:mm0}
\end{equation}
Indeed, it is unnatural if the mass of the monopole
depends on the detail of the UV theory.

Now I show that
the magnetic WGC,
in particular the bound on the UV cut-off scale
\eqref{eq:cutWGC}
is satisfied
in EFT$_{\mathrm{eff}}$,
provided additional assumptions, 
which I make explicit below, hold.
It is natural to assume that
the vacuum expectation value of the Higgs field
is below the UV cut-off scale of EFT$_{\mathrm{UV}}$,
$v < \Lambda \lesssim g M_P/\sqrt{2}$.
Then, it follows that
\begin{align}
\Lambda_{\mathrm{eff}} &\sim \frac{1}{L_m}
\sim
\frac{g v}{\sqrt{Z}}
\nn\\
&
\lesssim
\frac{g^2 M_P}{\sqrt{2Z}}
=
\left(\frac{g}{\sqrt{2Z} \sin \gamma} \right) 
g_{\mathrm{eff}} M_P = 
\frac{g Z}{\sqrt{2Z} \cos \gamma} 
g_{\mathrm{eff}} M_P
\nn\\
&\lesssim \frac{1}{\sqrt{Z}} g_{\mathrm{eff}} M_P 
\lesssim g_{\mathrm{eff}} M_P 
\, ,
\label{eq:magWGCeff}
\end{align}
i.e. the bound on the UV cut-off scale
\eqref{eq:cutWGC}
associated with $U(1)_{\mathrm{eff}}$ gauge theory
is (safely) satisfied.
In deriving \eqref{eq:magWGCeff}, 
I used 
$\sqrt{2} \cos\gamma \geq 1$ for
$Z \geq 1$, and
I assumed $g Z \lesssim 1$
so that the perturbative description
of EFT$_{\mathrm{UV}}$ is valid
(recall that the Higgs field had charge $(Z,1)$
in the original charge basis).

The two extra assumptions I made
to show \eqref{eq:magWGCeff}
were as follows:\footnote{%
These assumptions were also made in
the analysis by Saraswat \cite{Saraswat:2016eaz}.}
\begin{align}
\mbox{Assumption }1&:\quad v < \Lambda \, .\\
\mbox{Assumption }2&:\quad g Z \lesssim 1 \, .
\end{align}
To analyze the assumption 1,
I first assume a quadratic potential for the Higgs field:
\begin{equation}
V_H(H) 
= -\frac{\mu^2}{2} | H |^2 + \frac{\lambda}{4!} | H |^4 \, .
\label{eq:VH}
\end{equation}
Then, the vacuum expectation value $v$ is given by
\begin{equation}
v = \sqrt{\frac{6}{\lambda}} \, \mu \, .
\label{eq:v}
\end{equation}
Here, 
I chose the gauge
so that the vacuum expectation value $v$ becomes real.
I already assumed that $m_H = \sqrt{2}\mu < \Lambda$ 
in \eqref{eq:mAmB}
so that the 
creation of Higgs particles
can be described within the EFT$_{\mathrm{UV}}$.
Therefore, in order to achieve $v \gg \Lambda$,
a very small parameter
$\lambda \ll 1$ is required.
Thus in order to remove the assumption 1,
one need to introduce a tiny coupling
which in view of naturalness requires an explanation.
Also from naturalness, 
\eqref{eq:VH} is the most relevant part of the potential
for the Higgs field, and the above estimate of
the expectation value should not change significantly
by introducing higher order interactions.

The assumption 2 is a requirement that
EFT$_{\mathrm{UV}}$
is in the perturbative regime.
It would be difficult to make any quantitative conclusion
without this assumption.

Now, I examine whether the electric WGC bound
\eqref{eq:eWGC}
is respected by the $\psi_B$ particles.
The charge $\tilde{q}_B$ of the $\psi_B$ particles 
with respect to
the gauge group $U(1)_{\mathrm{eff}}$
in my convention is 
$\tilde{q}_B = Z$,
see Table \ref{table:charges}.
From \eqref{eq:geff}
it follows that
\begin{equation}
\frac{g_{\mathrm{eff}} \tilde{q}_B M_P}{m_B}
=
\frac{g \cos \gamma M_P}{m_B}
\gtrsim
\frac{\sqrt{2} \Lambda \cos \gamma}{m_B}  \gtrsim 1 \, ,
\label{eq:BeWGC}
\end{equation}
where the model assumptions for
EFT$_{\mathrm{UV}}$, i.e. 
\eqref{eq:mAmB}
and
\eqref{eq:LambdaUV}
and
$\sqrt{2} \cos \gamma \geq 1$
for 
$Z \geq 1$ have been used.
Thus $\psi_B$ particles 
satisfy the electric WGC bound \eqref{eq:eWGC}.
In fact,
one can show more generally
that 
the weak electric WGC will not be violated
by a spontaneous gauge symmetry breaking.
This is explained in Appendix~\ref{notviolated}.
However, the strong electric WGC can be violated
in EFT$_{\mathrm{eff}}$
as shown below.

As for the $\psi_A$ particles,
their charge 
with respect to $U(1)_{\mathrm{eff}}$
is one in my convention.
When
$m_A > \Lambda_{\mathrm{eff}}$
which is the case under study,
$\psi_A$ may violate the electric WGC bound
with respect to $U(1)_{\mathrm{eff}}$:
\begin{equation}
\frac{g_{\mathrm{eff}}M_P}{m_A} 
<
\frac{g \sin \gamma M_P}{\Lambda_{\mathrm{eff}}}
=
{g M_P \sin \gamma}\frac{\sqrt{Z}}{g v}
=
\frac{M_P}{v} \sqrt{Z} \sin \gamma \, ,
\label{eq:Aviolate}
\end{equation}
where \eqref{eq:Lambdaeff} has been used.
The vacuum expectation value of the Higgs field $v$
is an input parameter of EFT$_{\mathrm{UV}}$ 
which controls the gauge symmetry breaking scale.
For a given $v$,
the right hand side of 
\eqref{eq:Aviolate}
can be made arbitrary small
by taking $Z$ large.
Therefore, $\psi_A$ particles
may violate the electric WGC bound,
depending on the 
choice of the parameter $Z$.

When $m_A < m_B$,
the lightest particle charged with respect to 
$U(1)_{\mathrm{eff}}$ is $\psi_A$.
Therefore, when $\psi_A$ particles do not
satisfy the electric WGC bound \eqref{eq:eWGC},
the strong version of the electric WGC is violated in EFT$_\mathrm{eff}$,
while it was satisfied in EFT$_{\mathrm{UV}}$.
This is a notable feature of the Saraswat model \cite{Saraswat:2016eaz}.

\subsubsection*{Case 2: $\psi_A$ is known, 
$\psi_B$ is unknown to the low energy observers
associated with EFT$_{\mathrm{eff}}$}
As discussed in case 1 above, 
$\psi_A$ particles are not guaranteed to
satisfy the electric WGC bound 
with respect to the $U(1)_{\mathrm{eff}}$ gauge group.
On the other hand,
the field $\psi_B$ gives rise to particles
which satisfy the electric WGC bound \eqref{eq:eWGC}.
Therefore, if the low energy observers
associated with EFT$_{\mathrm{eff}}$
do not know the existence of $\psi_B$ particles,
when $\psi_A$ particles do not satisfy
the WGC bound with respect to the $U(1)_{\mathrm{eff}}$ gauge group,
the electric WGC may \textit{appear} to be violated
for the low energy observers 
associated with EFT$_{\mathrm{eff}}$.

\subsubsection*{Case 3: $\psi_A$ is unknown, 
$\psi_B$ is known to to the low energy observers
associated with EFT$_{\mathrm{eff}}$}

In this case, 
the charged particles known to 
the low energy observers associated with
the EFT$_{\mathrm{EFT}}$
are only $\psi_B$ particles.
Then, in my convention,
the gauge coupling constant is normalized as
\begin{equation}
g_{\mathrm{eff2}}  = g \cos \gamma 
= g_{\mathrm{eff}} \cot \gamma \, .
\label{eq:geff2}
\end{equation}
Here, I have introduced a new notation $g_{\mathrm{eff2}}$
to distinguish it from the 
coupling constant $g_{\mathrm{eff}}$ 
which was used in the case 1.
Then, the low energy observers associated with EFT$_{\mathrm{eff}}$
predict that
the bound on the UV cut-off scale $\Lambda_{\mathrm{eff}}$
of EFT$_{\mathrm{eff}}$ 
is given by
\begin{equation}
\Lambda_{\mathrm{eff}} \lesssim g_{\mathrm{eff2}} M_P 
= g_{\mathrm{eff}} \cot \gamma M_P \, .
\label{eq:bound2}
\end{equation}
Since 
$\cot \gamma \geq 1$ for $Z \geq 1$,
comparing with
\eqref{eq:magWGCeff}
it is clear that \eqref{eq:bound2} is satisfied.

\subsubsection*{Case 4: Neither $\psi_A$ nor $\psi_B$ 
are known to to the low energy observers
associated with EFT$_{\mathrm{eff}}$}

In this case,
there is no charged particle known to
the low energy observers associated with EFT$_{\mathrm{eff}}$.
Then, the electric WGC appears to be violated,
assuming that the low energy observers
know the existence of the gauge field through 
some non-minimal coupling to the gauge field.
Otherwise, the low energy observers
do not recognize EFT as gauge theory,
and there is no room to discuss the WGC from the beginning.
Without a charged particle
minimally coupled to the gauge field known,
the low energy observers do not know
what could be the smallest magnetic charge of the monopole.
The lack of the knowledge can be stated as the bound.
Without any information of the smallest charge,
the low energy observers can say 
$g q_0 \lesssim 1$ is still not forbidden,
assuming that the EFT is in perturbative regime,
where $q_0$ is the possible smallest charge.
In the current convention $q_0=1$, 
and this gives a bound $g\lesssim 1$.
It follows that the mass of the lightest monopole is bounded as 
$m_{m0} \gtrsim \Lambda$ as in \eqref{eq:mmono},
which gives the bound on the UV cut-off scale
$\Lambda \lesssim M_P$.
However, one usually assumes 
that the UV cut-off scale of EFTs
is below the Planck scale.
Thus the WGC 
does not add more constraint on
the UV cut-off scale of the EFT under consideration
than what one usually already assumes in EFTs.

\subsection{Implications to the bottom-up 
model buildings}\label{subsec:model}

Below I draw lessons from the analysis of the
Saraswat model
for general bottom-up model buildings.
In Sec.~\ref{subsec:BH},
after the examination of the BH discharge processes,
I assumed that the actual discharge process
need not be efficient.
Then, it followed that
the charged particles which satisfy
the electric WGC need not have mass
below the UV cut-off scale of EFTs.
If this is the case,
the constraining power of the electric WGC 
on EFTs
is quite limited.
In fact, usually in the model building,
one does not worry too much about stable heavy particles
whose mass is above the UV cut-off scale of 
the EFT.
This is because their quantum effects
are renormalized into the parameters
of the low energy EFT
and cannot be separately measurable
in the low energy experiments
\cite{Appelquist:1974tg,Ovrut:1980eq}.
Their classical effects as sources
can be taken into account,
but if these stable heavy particles are not around,
there is no need to include them.
What this means in the bottom-up approach to model buildings
is that
if one has a model which does not satisfy the electric WGC,
one can simply assume that such particle exists
with mass above the UV cut-off scale of the EFT
to make this model satisfy the electric WGC.
Theoretically, these stable heavy particles can be 
included in EFT as HPET,
but practically, as long as one does not have such particles around,
there would not be much need for studying processes
involving the stable heavy particles.

However, the bound on the UV cut-off scale
\eqref{eq:cutWGC} derived from the magnetic
WGC still gives an interesting constraint.
To state the magnetic WGC,
the low energy observers
associated with the EFT
need to know the existence of at least one charged particle state.
With respect to the known charge,
the low energy observers can give
physically meaningful normalization of
the gauge coupling constant,
which can be used to 
unambiguously state the magnetic WGC
and the bound on the UV cut-off scale of the EFT.
In this article, I use the normalization
with which
the smallest charge known to the low energy observers is one.
With this convention, 
the bound on the UV cut-off scale 
is stated as in \eqref{eq:cutWGC},
which
gives a non-trivial constraint on EFTs from the WGC.
As I have stressed, 
the knowledge of the low energy observers 
could be limited,
and as a consequence 
the bound on the UV cut-off scale could be
looser than 
the true bound derived from 
the magnetic WGC assuming it holds.
However, a bound looser than the true bound is not wrong,
it is just less constraining.
It still gives the best bound
for the given knowledge of the smallest charge.
The case 3 was an example
of the looser bound on the UV cut-off scale
predicted by the low energy observers.

After the general discussions above,
now
I examine the implications of 
the limitation of the low energy observers
associated with EFTs further
in a specific example:
The extra-natural inflation model
based on the Saraswat model will be studied.

\subsubsection*{Implications to extra-natural inflation 
models}

One of the original motivations of 
the WGC was to explain
why it appeared difficult to realize
extra-natural inflation in string theory
\cite{ArkaniHamed:2003wu}
compatible with the observations such as \cite{Ade:2015lrj}.
In natural inflation models,
the inflaton need to make a super-Planckian field excursion
in order to be consistent with the observations of
the spectral tilt and 
the bound on the tensor-to-scalar ratio,
assuming 
the number of e-folds to be $50-60$.
The extra-natural inflation realizes the super-Planckian
field excursion
at the cost of tiny gauge coupling constant
(typically $\Ord(10^{-3})$ or less
\cite{ArkaniHamed:2003wu,Furuuchi:2013ila})
and the non-local nature of Wilson loops
winding the extra dimension.
The magnetic WGC, or more explicitly 
the bound on the UV cut-off scale \eqref{eq:cutWGC},
poses an obstruction for realizing 
such tiny gauge coupling constant
with a required UV cut-off scale.

The Saraswat model discussed in the previous section
achieves a tiny gauge coupling constant $g_{\mathrm{eff}}$
from the large hierarchy between charges, $Z \gg 1$.
I examine 
whether this mechanism can help achieving 
the super-Planckian inflaton excursion.
It is straightforward to generalize
the Saraswat model discussed in the previous section
to $5$D.
Then, the $5$-th component of 
the $5$D gauge fields $A_\mu$ and $B_\mu$,
with $\mu$ now runs from $0,\cdots,3$ and $5$,
gives rise to axionic scalar fields
in $4$D with 
periodicity $2\pi f$,\footnote{The periodicity of the axion fields
is a consequence of the gauge symmetry, 
and it remains even after the gauge field acquires 
a mass \cite{Furuuchi:2015foh}.}
where the axion decay constant $f$ is given by
\begin{equation}
f = \frac{1}{2\pi g L_5} \, .
\label{eq:f}
\end{equation}
Here, $L_5$ is the radius of the compact $5$-th direction,
which I assume to be a circle.
The $4$D gauge coupling constant $g$
is related to the $5$D gauge coupling constant $g_5$
as $g = g_5/\sqrt{2\pi L_5}$.
I require that
the compactification scale
is below the UV cut-off scale $\Lambda$ of the EFT$_{\mathrm{UV}}$:
\begin{equation}
\frac{1}{L_5} \lesssim \Lambda  \lesssim \frac{g M_P}{\sqrt{2}} \, ,
\label{eq:L5}
\end{equation}
where I assumed that
the bound on the UV cut-off scale \eqref{eq:LambdaUV} holds.
Putting \eqref{eq:f} into \eqref{eq:L5},
a bound on the axion decay constant $f$ is obtained:
\begin{equation}
f \lesssim M_P \, .
\label{eq:fMP}
\end{equation}
The constraint on the axion decay constant like
\eqref{eq:fMP} poses an obstruction 
for achieving a super-Planckian
field excursion of inflaton in
the extra-natural inflation model with a single gauge field.
However, in the Saraswat model,
there are two gauge fields, 
and 
the inflaton trajectory need not be parallel
to the periodic direction.
As a consequence, the inflaton field excursion need
not be restricted by the period $2\pi f$.
Indeed, the spontaneous gauge symmetry breaking
adds a potential
which tilt the inflaton trajectory from
the periodic direction.
I denote the canonically normalized
fields which arise from
the zero-modes of the
$5$-th dimensional component of 
the gauge fields $A_5$ and $B_5$
as ${A}$ and ${B}$,
respectively.
Fields with charge vectors
$(1,0)$ and $(0,1)$
whose masses are
light compared with the compactification scale $L_5$
contribute to the potential 
for $A$ and $B$
(see for example
\cite{Furuuchi:2014cwa}
for the details of the calculations).
Since I assumed $1/L_5 < \Lambda$ as in \eqref{eq:L5},
following the discussions in
Sect.~\ref{subsec:trivial},
these light charged fields
give rise to particles
which satisfy the electric WGC bound \eqref{eq:eWGC}
with respect to the gauge group $U(1)_A \times U(1)_B$.
Thus in extra-natural inflation models,
the particles which satisfy the electric WGC bound 
are included in order to generate a suitable inflaton potential.
The potential for the field $A$ and the field $B$ is given as
\begin{align}
V(\tilde{A},\tilde{B})
&\sim
C_1 \cos \left( \frac{A}{f} \right)
+
C_2 \cos \left( \frac{B}{f} \right)
+
\frac{1}{2}
\left(
g Z v A + g v B
\right)^2 
\nn\\
&\sim
C_1 \cos \left( \frac{\cos \gamma \tilde{A} - \sin \gamma \tilde{B}}{f} \right)
+
C_2 \cos \left( \frac{\sin \gamma \tilde{A} + \cos \gamma \tilde{B}}{f} \right)
+
\frac{m_V^2}{2}\tilde{A}^2 \, .
\label{eq:pot}
\end{align}
Here,
$\tilde{A}$ and
$\tilde{B}$ 
are defined using 
the linear transformation 
as in \eqref{eq:rot}:
\begin{align}
\left(
\begin{array}{c}
\tilde{A}\\
\tilde{B}
\end{array}
\right)
&=
\left(
\begin{array}{cc}
\cos \gamma & \sin \gamma \\
-\sin \gamma & \cos \gamma
\end{array}
\right)
\left(
\begin{array}{c}
{A}\\
{B}
\end{array}
\right) \, ,
\label{eq:infrot}
\end{align}
where
\begin{equation}
\cos \gamma := \frac{Z}{\sqrt{1+Z^2}} \, ,
\quad
\sin \gamma := \frac{1}{\sqrt{1+Z^2}}\, .
\label{eq:gamma2}
\end{equation}
In \eqref{eq:pot},
$m_V = g \sqrt{1+Z^2} \, v$ as
in \eqref{eq:NPS1}, and
$C_1$ and $C_2$ depend on the number of
the light fields with charge vector
$(1,0)$ and $(0,1)$, respectively.
The field $\tilde{B}$ is to be identified
with the inflaton.
Then, 
to achieve the large axion decay constant,
I assume
$C_1 \gg C_2$
so that the second sinusoidal potential 
is suppressed.\footnote{%
The periodic modulation by the second sinusoidal potential
is subject to the observational constraints
\cite{Flauger:2009ab}.}
I further assume
$C_1/f \ll m_V^2 \tilde{A}$ so that the field
$\tilde{A}$ settles down to its minimum 
$\tilde{A} \sim 0$ first.
Under these assumptions, 
I obtain the effective potential for the field 
$\tilde{B}$ as
\begin{equation}
V_{\mathrm{eff}}(\tilde{B}) 
\sim 
C_1 
\cos 
\left( \frac{-\tilde{B}}{f_{\mathrm{eff}}} \right) \, ,
\label{eq:Vinf}
\end{equation}
where
\begin{equation}
f_{\mathrm{eff}} = \frac{f}{\sin \gamma} 
\lesssim \frac{M_P}{\sin \gamma} \, .
\label{eq:feff}
\end{equation}
Therefore, the bound on the effective axion decay constant 
$f_{\mathrm{eff}}$ is much milder than that for $f$ \eqref{eq:fMP}
if $Z \gg 1$,
and $f_{\mathrm{eff}}$ can be super-Planckian if $Z$ is large enough.
This 
enhancement of the effective axion decay constant
by tilting of the inflaton trajectory with respect
to the periodic direction is
the same mechanism employed in
axion-monodromy inflation models
\cite{Kim:2004rp,Silverstein:2008sg,McAllister:2008hb,Berg:2009tg,%
Kappl:2014lra,Ben-Dayan:2014zsa}.
However, in the current model,
the monodromy 
(going round in periodic direction
more than once)
itself is not used.

Thus the extra-natural inflation model
based on the Saraswat model
may achieve super-Planckian inflaton travel
without violating the WGC.
However,
the large hierarchy between charges
$Z \gg 1$ required to achieve
the super-Planckian inflaton excursion
is a potential obstruction for embedding this model to string theory
\cite{Ibanez:2017vfl,Shiu:2013wxa,McAllister:2016vzi,Hebecker:2017lxm}.

It is important to notice that
the extra-dimensional aspect of this model 
can only be described by the high energy theory
EFT$_{\mathrm{UV}}$, not EFT$_{\mathrm{eff}}$.
This is because when 
the super-Planckian effective axion decay constant
$f_{\mathrm{eff}} \gtrsim M_P$
is achieved, from \eqref{eq:magWGCeff}
\begin{equation}
\Lambda_{\mathrm{eff}}
\lesssim
g_{\mathrm{eff}} M_P
\lesssim
g_{\mathrm{eff}} f_{\mathrm{eff}}
\sim \frac{1}{L_5} \, .
\label{eq:noteff}
\end{equation}
Here, I have used
\begin{equation}
g_{\mathrm{eff}} f_{\mathrm{eff}}
=
g f
=
\frac{1}{2\pi L_5} \, ,
\label{eq:gfgf}
\end{equation}
which follows from \eqref{eq:f}, \eqref{eq:feff}
and \eqref{eq:geff}.
Thus the low energy observers
associated with EFT$_{\mathrm{eff}}$
cannot resolve the extra dimension.
This means that
while
EFT$_{\mathrm{eff}}$ achieves a tiny gauge coupling constant,
it cannot be used as a model of extra-natural inflation.
It is a genuinely $4$D model
and cannot \textit{explain} 
the super-Planckian axion decay constant.
The $5$D model EFT$_{\mathrm{UV}}$, 
which satisfies  
\eqref{eq:L5}, 
is required as a model of 
extra-natural inflation.
Accordingly, the relevant WGC 
relevant for constraining
the inflaton field range
is that on EFT$_{\mathrm{UV}}$,
not the WGC on EFT$_{\mathrm{UV}}$.
In EFT$_{\mathrm{UV}}$
the gauge group is $U(1)_A \times U(1)_B$
with the gauge coupling constant $g$.
Thus it is more appropriate to view
the extra-natural inflation model
based on the $5$D version of the Saraswat model
as a model achieving the super-Planckian
inflaton travel by the tilt 
of the inflaton trajectory from the periodic direction,
rather than to view it as an extra-natural inflation model
with a tiny gauge coupling constant $g_{\mathrm{eff}}$.

\section{Summary and discussions}\label{sec:discussions}

In this article,
I examined the WGC from the low energy observers'
perspective,
to address the issue of 
to what extent the WGC actual constrains EFTs.
For this purpose, I introduced idealized
low energy observers for a given EFT,
who have full access to particles
whose mass is
below the UV cut-off scale $\Lambda$
of the EFT.
However, 
I did not assume that 
all the stable particles whose mass 
is above $\Lambda$ 
are known to the low energy observers.
This is because
creations of particles heavier than $\Lambda$ 
occur only
at energy higher than $\Lambda$
and those are not under control of the EFT.
Since I considered idealized low energy observers,
this limitation of the knowledge of the low energy observers
is fundamental and it is
not due to experimental practicalities.
An immediate consequence
of the limitation of the knowledge 
of the low energy observers
associated with 
Abelian gauge EFTs
was that
the low energy observers 
can never be sure what is the smallest charge in Nature.
It also follows that
they are not sure about the smallest magnetic charge neither.
What the low energy observers can do
was to constrain possible magnetic charges
based on the knowledge of the smallest electric charge
via the Dirac quantization condition.
The existence of a charged particle
was necessary to give a physically meaningful
normalization of the gauge coupling constant,
and it was crucial for unambiguously stating
the magnetic WGC.

I gave two main reasons
for considering stable heavy particles
whose mass is above the UV cut-off scale
of the EFT under consideration.
The first reason was that
it is of practical importance to ask
if the low energy IR EFT satisfies the WGC
when it is derived from UV EFT which
satisfies the WGC.
The UV EFT may have charged particles whose mass is
below its UV cut-off scale,
but their mass can be above the 
UV cut-off scale of the IR EFT.
If this is the case,
it is easy to construct a model
in which in the UV EFT the WGC is satisfied
but in the IR EFT the WGC appears to be violated,
if the low energy observers 
associated with the IR EFT know nothing about
the stable heavy particles whose mass is above
the UV cut-off scale of the IR EFT.
When the WGC appears to be violated,
it does not constrain the IR EFT.
The WGC gives non-trivial constraints only when
the low energy observers associated with the IR EFT
know all or some of the stable heavy particles
whose mass is above the UV cut-off scale of the IR EFT.
The examination of the BH discharge arguments
in Sec.~\ref{subsec:BH} enhances this reasoning
for considering stable particles above 
the UV cut-off scale of the EFT under consideration.
The charged particles responsible
for the discharge of BHs can have 
non-zero mass.
This means that
one can integrate out degrees of freedom
above the mass of the 
lightest charged particle
which satisfies the electric WGC bound
\eqref{eq:eWGC} to obtain an EFT.
Since there is no charged particle in the EFT thus obtained,
the electric WGC appears to be violated
unless one takes into account the 
charged particles whose mass is above
the UV cut-off scale of the EFT.
The second reason for
taking into account 
stable heavy particles
was that
once the magnetic WGC,
or more precisely,
the bound on the UV cut-off scale 
\eqref{eq:cutWGC}
is assumed,
charged particles whose mass is below
the UV cut-off scale automatically
satisfy the electric WGC bound,
as explained in Sec.~\ref{subsec:trivial}.
Therefore, the electric WGC
becomes a non-trivial question
only when there is no charged particle
below the UV cut-off scale.

I illustrated the above points
taking the Saraswat model \cite{Saraswat:2016eaz}
as an example.
In the Saraswat model,
strong electric WGC can be violated at low energy,
while it is respected at high energy.
When some of the charged particles
whose mass is above the UV cut-off scale of the low energy EFT
are not known to the low energy observers,
the weak electric WGC may also
appear to be violated for
the low energy observers.
In this article, 
I also showed that
the magnetic WGC is not violated at low energy.
To show this result, it was important
to correctly identify
the UV cut-off scale of the low energy theory.
This was done by
a detailed analysis of the internal structure
of the monopole of $U(1)_{\mathrm{eff}}$.
the fractional magnetic charges 
of the constituent monopoles
with respect to $U(1)_{\mathrm{eff}}$
reveals the break-down of the low energy description.
The 
$U(1)_{\mathrm{broken}}$ gauge field
enters as
the new physics which replaces the low energy description.
While the massive gauge bosons still do not appear as particle states
at the energy scale corresponding to 
the size of the monopole of $U(1)_{\mathrm{eff}}$,
their existence is required
to keep the Dirac strings unphysical.

The limitation of the knowledge of 
the low energy observers
on the existence of charged particles
whose mass is above the UV cut-off scale of the EFT
leads to the limitation of the constraining
power of the WGC on bottom-up
model buildings using EFTs.
Regarding the electric WGC,
one can freely construct a model based on an EFT,
and if the model does not contain a particle
which satisfies the electric WGC bound \eqref{eq:eWGC},
one can simply assume the existence of
particles which satisfy the electric WGC bound
above the UV cut-off scale.
One can further assume that 
they were not created with observable density.
Since below the UV cut-off scale
the quantum effects of the stable heavy particles
are renormalized into parameters of the EFT,
the additional assumption does not 
practically modify the model.
On the other hand, 
when a charged particle is known
to the low energy observers,
one can give a physical definition
of the gauge coupling constant to state
the magnetic WGC which in turn puts a bound
on the UV cut-off scale of the EFT.
Due to the limitation of the knowledge of 
the low energy observers
regarding the charges of particles,
the bound on the UV cut-off scale can be looser
than the one obtained from the true smallest charge
assuming the magnetic WGC to hold.
Nevertheless, the looser bound is not violated
by the true bound,
it is just less constraining.
But the looser bound is still the best bound
for the given knowledge.

The lessons from the analysis of the Saraswat model
for general bottom-up model buildings based on EFTs 
were discussed
in Sec.~\ref{subsec:model}.
The implications
to extra-natural inflation models
were studied in some detail.
The extra-natural inflation model
based on the Saraswat model provides an example
in which WGC alone does not forbid 
super-Planckian inflaton excursion.
However, in this model,
the large hierarchy between charges
required to achieve the super-Planckian inflaton excursion
can be a potential obstacle for realizing 
this models in string theory.
At the moment,
there is no universal formula
for the bound on the hierarchy between charges
which applies to a large class of EFTs.
It will be interesting to
look for a criterion
for separating out EFTs
which cannot be realized in string theory
due to the large hierarchy between charges.

As I discussed in Sec.~\ref{subsec:BH},
the requirement that
BH should be able to release
more charges than mass in Planck unit
does not constrain the mass of the charged particles,
under the assumption I made that
BHs only need to be kinematically allowed to discharge, but
the actual physical process for the discharge need not be efficient. 
In particular, the mass of the charged particle
which satisfy the electric WGC bound
can be above the UV cut-off scale
of the EFT under consideration.
The lack of necessity of the efficient BH discharge process
made me suspect that 
the BH discharge process
may not be relevant for realizing
the low energy world satisfying the WGC.
I suspect that there is a more fundamental principle
which prevents the WGC to be violated.
A possible clue for identifying 
such hypothetical fundamental principle
may be that
for the magnetic WGC, there is a
physically independent
argument
in support for the bound on the 
UV cut-off scale:
The monopole with the smallest magnetic charge 
should not be a BH
\cite{ArkaniHamed:2006dz}.
Indeed, it is strange if the monopole
with the smallest magnetic charge 
is already a BH,
given a widely accepted belief that
the Bekenstein-Hawking entropy
has a microscopic origin.
I suspect that 
this argument may be
more
directly related to the fundamental principle behind the WGC,
which is likely to be some sort of 
entropy bound.\footnote{%
Related ideas have been pursued in 
\cite{delaFuente:2014aca,Fisher:2017dbc}.}
However, heuristic arguments based on an entropy bound
should be taken with caution, 
in light of the quantitative understanding of 
the Bekenstein bound \cite{Casini:2008cr}.
In the context of giving more emphasis on
the magnetic WGC and the bound on the 
UV cut-off scale,
I find it indicative that
once the magnetic WGC, in particular the
bound on the UV cut-off scale on an EFT is assumed,
the electric WGC automatically follows
if there exists a charged particle
whose mass is below the UV cut-off scale.
This observation may provide a step
for deriving the electric WGC from
the hypothetical entropy bound.\\

\begin{center}
\textbf{Acknowledgments}\\
\end{center}
I would like to thank Yoji Koyama
for collaboration at the early stage of this work
as well as careful reading of the manuscript.
I would also like to thank Cheng-Yang Lee for useful discussions.
This work is supported in part by 
the Science and Engineering Research Board,
Department of Science and Technology, Government of India
under the grant No.~EMR/2015/002471.

\appendix
\section{The weak electric WGC is not violated by 
spontaneous gauge symmetry breakings}\label{notviolated}

In this appendix, I show that
starting from gauged EFT with
product of $U(1)$ gauge groups,
the weak electric WGC will never be violated
by spontaneous gauge symmetry breaking.
For this purpose,
I need to recall the result of
\cite{Cheung:2014vva}.
Let
$\Pi_{\alpha=1}^N U(1)_\alpha$
be the gauge group of some EFT.
In this appendix, I use a normalization
for the gauge coupling and the charge
different 
from the main body:
The product of the charge and the gauge coupling
$g_\alpha q_\alpha$ (no sum in $\alpha$)
in the notation of the main body
is simply denoted as $q_\alpha$ in this appendix.

Consider a BH with charge $\vec{Q}$
and mass $M$.
I define
charge-to-mass vector $\vec{Z}$ as
\begin{equation}
\vec{Z} = \frac{\vec{Q} M_P}{M} \, .
\label{eq:vecZ}
\end{equation}
Consider a
decay of the BH
to a final state
consisting of
$n_i$ particles of spices $i$
with charge vector $\vec{q}_i$ and mass $m_i$.
The charge conservation implies
\begin{equation}
\vec{Q} = \sum_i n_i \vec{q}_i \, .
\label{eq:vecQ}
\end{equation}
The energy conservation implies
\begin{equation}
M > \sum_i n_i m_i \, .
\label{eq:Econserve}
\end{equation}
Let $\sigma_i := n_i m_i /M$
be the spices $i$
fraction of the total final state mass.
Then \eqref{eq:vecQ} can be rewritten as
\begin{equation}
\vec{Z} = \sum_i \sigma_i \vec{z}_i \, ,
\label{eq:Q2}
\end{equation}
where 
\begin{equation}
\vec{z}_i := \frac{\vec{q}_i M_P}{m_i} \, ,
\label{eq:zi}
\end{equation}
while \eqref{eq:Econserve} can be rewritten as
\begin{equation}
1 > \sum_i \sigma_i  \, .
\label{eq:E2}
\end{equation}
Thus the requirement that the BH is
kinematically allowed to decay
amounts to the condition
that $\vec{Z}$ be a sub-unitary weighted
average of $\vec{z}_i$.
The geometric interpretation of this condition
is that
the $N$-dimensional 
unit ball  $|\vec{Z}| \leq 1$
representing the BPS bound
\eqref{eq:BPS}
is contained in
the convex hull spanned
by the vectors $\vec{z}_i$.

In this article,
I focused on Coulomb force
mediated by massless gauge fields.
In this setting,
a spontaneous gauge symmetry breaking
%
amounts to 
projecting charge vectors
into space spanned by
charges of the remaining massless $U(1)$'s.\footnote{%
The charge of BHs associated with massive gauge fields
may be probed using Aharonov-Bohm effect \cite{Preskill:1990ty}.
Implication of this observation
to the WGC is currently not clear to me, 
but it may be an interesting direction to investigate.}
Suppose that $p < N$ independent linear combinations
of the $U(1)$ gauge fields become massive.
I denote the projection
$P$ to the subspace in the charge-to-mass vector space
where 
charges for the massless gauge fields remain. 
Clearly,
if the $N$-dimensional unit ball is contained in 
the original convex hull
spanned by the vectors $\vec{z}_i$,
the $(N-p)$-dimensional unit ball in the projected space is also
contained in the convex hull spanned by the projected
vectors $P \vec{z}_i$.
This means that if one starts from the EFT which respects
the weak electric WGC,
the weak electric WGC will not be violated
by the spontaneous gauge symmetry breaking.

\bibliography{WGCRef}

\providecommand{\href}[2]{#2}\begingroup\raggedright\begin{thebibliography}{10}

\bibitem{Saraswat:2016eaz}
P.~Saraswat, ``{Weak gravity conjecture and effective field theory},''
  \href{http://dx.doi.org/10.1103/PhysRevD.95.025013}{{\em Phys. Rev.}
  {\bfseries D95} no.~2, (2017) 025013},
\href{http://arxiv.org/abs/1608.06951}{{\ttfamily arXiv:1608.06951 [hep-th]}}.

\bibitem{Vafa:2005ui}
C.~Vafa, ``{The String landscape and the swampland},''
\href{http://arxiv.org/abs/hep-th/0509212}{{\ttfamily arXiv:hep-th/0509212
  [hep-th]}}.

\bibitem{ArkaniHamed:2006dz}
N.~Arkani-Hamed, L.~Motl, A.~Nicolis, and C.~Vafa, ``{The String landscape,
  black holes and gravity as the weakest force},''
  \href{http://dx.doi.org/10.1088/1126-6708/2007/06/060}{{\em JHEP} {\bfseries
  06} (2007) 060},
\href{http://arxiv.org/abs/hep-th/0601001}{{\ttfamily arXiv:hep-th/0601001
  [hep-th]}}.

\bibitem{Banks:2006mm}
T.~Banks, M.~Johnson, and A.~Shomer, ``{A Note on Gauge Theories Coupled to
  Gravity},'' \href{http://dx.doi.org/10.1088/1126-6708/2006/09/049}{{\em JHEP}
  {\bfseries 09} (2006) 049},
\href{http://arxiv.org/abs/hep-th/0606277}{{\ttfamily arXiv:hep-th/0606277
  [hep-th]}}.

\bibitem{ArkaniHamed:2003wu}
N.~Arkani-Hamed, H.-C. Cheng, P.~Creminelli, and L.~Randall, ``{Extra natural
  inflation},'' \href{http://dx.doi.org/10.1103/PhysRevLett.90.221302}{{\em
  Phys. Rev. Lett.} {\bfseries 90} (2003) 221302},
\href{http://arxiv.org/abs/hep-th/0301218}{{\ttfamily arXiv:hep-th/0301218
  [hep-th]}}.

\bibitem{Neubert:1996wg}
M.~Neubert, ``{Heavy quark effective theory},'' {\em Subnucl. Ser.} {\bfseries
  34} (1997) 98--165,
\href{http://arxiv.org/abs/hep-ph/9610266}{{\ttfamily arXiv:hep-ph/9610266
  [hep-ph]}}.

\bibitem{Manohar:2000dt}
A.~V. Manohar and M.~B. Wise, ``{Heavy quark physics},''
{\em Camb. Monogr. Part. Phys. Nucl. Phys. Cosmol.} {\bfseries 10} (2000)
  1--191.

\bibitem{Grozin:2004yc}
A.~G. Grozin, ``{Heavy quark effective theory},''
\href{http://dx.doi.org/10.1007/b79301}{{\em Springer Tracts Mod. Phys.}
  {\bfseries 201} (2004) 1--213}.

\bibitem{Guth:1980zm}
A.~H. Guth, ``{The Inflationary Universe: A Possible Solution to the Horizon
  and Flatness Problems},''
\href{http://dx.doi.org/10.1103/PhysRevD.23.347}{{\em Phys. Rev.} {\bfseries
  D23} (1981) 347--356}.

\bibitem{Einhorn:1980ik}
M.~B. Einhorn and K.~Sato, ``{Monopole Production in the Very Early Universe in
  a First Order Phase Transition},''
\href{http://dx.doi.org/10.1016/0550-3213(81)90057-2}{{\em Nucl. Phys.}
  {\bfseries B180} (1981) 385--404}.

\bibitem{Linde:1981mu}
A.~D. Linde, ``{A New Inflationary Universe Scenario: A Possible Solution of
  the Horizon, Flatness, Homogeneity, Isotropy and Primordial Monopole
  Problems},''
\href{http://dx.doi.org/10.1016/0370-2693(82)91219-9}{{\em Phys. Lett.}
  {\bfseries 108B} (1982) 389--393}.

\bibitem{Georgi:1990ak}
H.~Georgi and M.~B. Wise, ``{Superflavor Symmetry for Heavy Particles},''
\href{http://dx.doi.org/10.1016/0370-2693(90)90851-V}{{\em Phys. Lett.}
  {\bfseries B243} (1990) 279--283}.

\bibitem{Carone:1990pv}
C.~D. Carone, ``{Superflavor symmetry for heavy vector mesons and heavy
  quarks},''
\href{http://dx.doi.org/10.1016/0370-2693(91)91741-D}{{\em Phys. Lett.}
  {\bfseries B253} (1991) 408--410}.

\bibitem{Isgur:1989vq}
N.~Isgur and M.~B. Wise, ``{Weak Decays of Heavy Mesons in the Static Quark
  Approximation},''
\href{http://dx.doi.org/10.1016/0370-2693(89)90566-2}{{\em Phys. Lett.}
  {\bfseries B232} (1989) 113--117}.

\bibitem{Isgur:1989ed}
N.~Isgur and M.~B. Wise, ``{WEAK TRANSITION FORM-FACTORS BETWEEN HEAVY
  MESONS},''
\href{http://dx.doi.org/10.1016/0370-2693(90)91219-2}{{\em Phys. Lett.}
  {\bfseries B237} (1990) 527--530}.

\bibitem{Eichten:1989zv}
E.~Eichten and B.~R. Hill, ``{An Effective Field Theory for the Calculation of
  Matrix Elements Involving Heavy Quarks},''
\href{http://dx.doi.org/10.1016/0370-2693(90)92049-O}{{\em Phys. Lett.}
  {\bfseries B234} (1990) 511--516}.

\bibitem{Georgi:1990um}
H.~Georgi, ``{An Effective Field Theory for Heavy Quarks at Low-energies},''
\href{http://dx.doi.org/10.1016/0370-2693(90)91128-X}{{\em Phys. Lett.}
  {\bfseries B240} (1990) 447--450}.

\bibitem{Cheung:2014vva}
C.~Cheung and G.~N. Remmen, ``{Naturalness and the Weak Gravity Conjecture},''
  \href{http://dx.doi.org/10.1103/PhysRevLett.113.051601}{{\em Phys. Rev.
  Lett.} {\bfseries 113} (2014) 051601},
\href{http://arxiv.org/abs/1402.2287}{{\ttfamily arXiv:1402.2287 [hep-ph]}}.

\bibitem{Heidenreich:2015wga}
B.~Heidenreich, M.~Reece, and T.~Rudelius, ``{Weak Gravity Strongly Constrains
  Large-Field Axion Inflation},''
  \href{http://dx.doi.org/10.1007/JHEP12(2015)108}{{\em JHEP} {\bfseries 12}
  (2015) 108},
\href{http://arxiv.org/abs/1506.03447}{{\ttfamily arXiv:1506.03447 [hep-th]}}.

\bibitem{Heidenreich:2015nta}
B.~Heidenreich, M.~Reece, and T.~Rudelius, ``{Sharpening the Weak Gravity
  Conjecture with Dimensional Reduction},''
  \href{http://dx.doi.org/10.1007/JHEP02(2016)140}{{\em JHEP} {\bfseries 02}
  (2016) 140},
\href{http://arxiv.org/abs/1509.06374}{{\ttfamily arXiv:1509.06374 [hep-th]}}.

\bibitem{Hebecker:2015zss}
A.~Hebecker, F.~Rompineve, and A.~Westphal, ``{Axion Monodromy and the Weak
  Gravity Conjecture},'' \href{http://dx.doi.org/10.1007/JHEP04(2016)157}{{\em
  JHEP} {\bfseries 04} (2016) 157},
\href{http://arxiv.org/abs/1512.03768}{{\ttfamily arXiv:1512.03768 [hep-th]}}.

\bibitem{Heidenreich:2016aqi}
B.~Heidenreich, M.~Reece, and T.~Rudelius, ``{Evidence for a sublattice weak
  gravity conjecture},'' \href{http://dx.doi.org/10.1007/JHEP08(2017)025}{{\em
  JHEP} {\bfseries 08} (2017) 025},
\href{http://arxiv.org/abs/1606.08437}{{\ttfamily arXiv:1606.08437 [hep-th]}}.

\bibitem{Montero:2016tif}
M.~Montero, G.~Shiu, and P.~Soler, ``{The Weak Gravity Conjecture in three
  dimensions},'' \href{http://dx.doi.org/10.1007/JHEP10(2016)159}{{\em JHEP}
  {\bfseries 10} (2016) 159},
\href{http://arxiv.org/abs/1606.08438}{{\ttfamily arXiv:1606.08438 [hep-th]}}.

\bibitem{Gibbons:1975kk}
G.~W. Gibbons, ``{Vacuum Polarization and the Spontaneous Loss of Charge by
  Black Holes},''
\href{http://dx.doi.org/10.1007/BF01609829}{{\em Commun. Math. Phys.}
  {\bfseries 44} (1975) 245--264}.

\bibitem{Giudice:2008bi}
G.~F. Giudice, ``{Naturally Speaking: The Naturalness Criterion and Physics at
  the LHC},''
\href{http://arxiv.org/abs/0801.2562}{{\ttfamily arXiv:0801.2562 [hep-ph]}}.

\bibitem{Crisford:2017gsb}
T.~Crisford, G.~T. Horowitz, and J.~E. Santos, ``{Testing the Weak Gravity -
  Cosmic Censorship Connection},''
\href{http://arxiv.org/abs/1709.07880}{{\ttfamily arXiv:1709.07880 [hep-th]}}.

\bibitem{Parikh:1999mf}
M.~K. Parikh and F.~Wilczek, ``{Hawking radiation as tunneling},''
  \href{http://dx.doi.org/10.1103/PhysRevLett.85.5042}{{\em Phys. Rev. Lett.}
  {\bfseries 85} (2000) 5042--5045},
\href{http://arxiv.org/abs/hep-th/9907001}{{\ttfamily arXiv:hep-th/9907001
  [hep-th]}}.

\bibitem{Brezin:1970xf}
E.~Brezin and C.~Itzykson, ``{Pair production in vacuum by an alternating
  field},''
\href{http://dx.doi.org/10.1103/PhysRevD.2.1191}{{\em Phys. Rev.} {\bfseries
  D2} (1970) 1191--1199}.

\bibitem{Giddings:1995gd}
S.~B. Giddings, ``{The Black hole information paradox},'' in {\em {Particles,
  strings and cosmology. Proceedings, 19th Johns Hopkins Workshop and 5th
  PASCOS Interdisciplinary Symposium, Baltimore, USA, March 22-25, 1995}},
  pp.~415--428.
\newblock 1995.
\newblock
\href{http://arxiv.org/abs/hep-th/9508151}{{\ttfamily arXiv:hep-th/9508151
  [hep-th]}}.
\newblock

\bibitem{Hebecker:2015rya}
A.~Hebecker, P.~Mangat, F.~Rompineve, and L.~T. Witkowski, ``{Winding out of
  the Swamp: Evading the Weak Gravity Conjecture with F-term Winding
  Inflation?},'' \href{http://dx.doi.org/10.1016/j.physletb.2015.07.026}{{\em
  Phys. Lett.} {\bfseries B748} (2015) 455--462},
\href{http://arxiv.org/abs/1503.07912}{{\ttfamily arXiv:1503.07912 [hep-th]}}.

\bibitem{Ibanez:2017vfl}
L.~E. Ibanez and M.~Montero, ``{A Note on the WGC, Effective Field Theory and
  Clockwork within String Theory},''
\href{http://arxiv.org/abs/1709.02392}{{\ttfamily arXiv:1709.02392 [hep-th]}}.

\bibitem{Nielsen:1973cs}
H.~B. Nielsen and P.~Olesen, ``{Vortex Line Models for Dual Strings},''
\href{http://dx.doi.org/10.1016/0550-3213(73)90350-7}{{\em Nucl. Phys.}
  {\bfseries B61} (1973) 45--61}.

\bibitem{Nambu:1974zg}
Y.~Nambu, ``{Strings, Monopoles and Gauge Fields},''
\href{http://dx.doi.org/10.1103/PhysRevD.10.4262}{{\em Phys. Rev.} {\bfseries
  D10} (1974) 4262}.

\bibitem{Weinberg:1992hc}
E.~J. Weinberg, ``{Classical solutions in quantum field theories},''
\href{http://dx.doi.org/10.1146/annurev.ns.42.120192.001141}{{\em Ann. Rev.
  Nucl. Part. Sci.} {\bfseries 42} (1992) 177--210}.

\bibitem{Creutz:1974vk}
M.~Creutz, ``{The Higgs Mechanism and Quark Confinement},''
\href{http://dx.doi.org/10.1103/PhysRevD.10.2696}{{\em Phys. Rev.} {\bfseries
  D10} (1974) 2696}.

\bibitem{Ripka:2003vv}
G.~Ripka, ``{Dual superconductor models of color confinement},''
  \href{http://dx.doi.org/10.1007/b94800}{{\em Lect. Notes Phys.} {\bfseries
  639} (2004) pp.1--135},
\href{http://arxiv.org/abs/hep-ph/0310102}{{\ttfamily arXiv:hep-ph/0310102
  [hep-ph]}}.

\bibitem{Appelquist:1974tg}
T.~Appelquist and J.~Carazzone, ``{Infrared Singularities and Massive
  Fields},''
\href{http://dx.doi.org/10.1103/PhysRevD.11.2856}{{\em Phys. Rev.} {\bfseries
  D11} (1975) 2856}.

\bibitem{Ovrut:1980eq}
B.~A. Ovrut and H.~J. Schnitzer, ``{Decoupling Theorems for Effective Field
  Theories},''
\href{http://dx.doi.org/10.1103/PhysRevD.22.2518}{{\em Phys. Rev.} {\bfseries
  D22} (1980) 2518}.

\bibitem{Ade:2015lrj}
{\bfseries Planck} Collaboration, P.~A.~R. Ade {\em et~al.}, ``{Planck 2015
  results. XX. Constraints on inflation},''
  \href{http://dx.doi.org/10.1051/0004-6361/201525898}{{\em Astron. Astrophys.}
  {\bfseries 594} (2016) A20},
\href{http://arxiv.org/abs/1502.02114}{{\ttfamily arXiv:1502.02114
  [astro-ph.CO]}}.

\bibitem{Furuuchi:2013ila}
K.~Furuuchi and J.~M.~S. Wu, ``{$U(1)_{B-L}$ extra-natural inflation with
  Standard Model on a brane},''
  \href{http://dx.doi.org/10.1016/j.physletb.2013.12.054}{{\em Phys. Lett.}
  {\bfseries B729} (2014) 56--61},
\href{http://arxiv.org/abs/1310.4646}{{\ttfamily arXiv:1310.4646 [hep-ph]}}.

\bibitem{Furuuchi:2015foh}
K.~Furuuchi, ``{Excursions through KK modes},''
  \href{http://dx.doi.org/10.1088/1475-7516/2016/07/008}{{\em JCAP} {\bfseries
  1607} no.~07, (2016) 008},
\href{http://arxiv.org/abs/1512.04684}{{\ttfamily arXiv:1512.04684 [hep-th]}}.

\bibitem{Furuuchi:2014cwa}
K.~Furuuchi and Y.~Koyama, ``{Large field inflation models from
  higher-dimensional gauge theories},''
  \href{http://dx.doi.org/10.1088/1475-7516/2015/02/031}{{\em JCAP} {\bfseries
  1502} no.~02, (2015) 031},
\href{http://arxiv.org/abs/1407.1951}{{\ttfamily arXiv:1407.1951 [hep-th]}}.

\bibitem{Flauger:2009ab}
R.~Flauger, L.~McAllister, E.~Pajer, A.~Westphal, and G.~Xu, ``{Oscillations in
  the CMB from Axion Monodromy Inflation},''
  \href{http://dx.doi.org/10.1088/1475-7516/2010/06/009}{{\em JCAP} {\bfseries
  1006} (2010) 009},
\href{http://arxiv.org/abs/0907.2916}{{\ttfamily arXiv:0907.2916 [hep-th]}}.

\bibitem{Kim:2004rp}
J.~E. Kim, H.~P. Nilles, and M.~Peloso, ``{Completing natural inflation},''
  \href{http://dx.doi.org/10.1088/1475-7516/2005/01/005}{{\em JCAP} {\bfseries
  0501} (2005) 005},
\href{http://arxiv.org/abs/hep-ph/0409138}{{\ttfamily arXiv:hep-ph/0409138
  [hep-ph]}}.

\bibitem{Silverstein:2008sg}
E.~Silverstein and A.~Westphal, ``{Monodromy in the CMB: Gravity Waves and
  String Inflation},'' \href{http://dx.doi.org/10.1103/PhysRevD.78.106003}{{\em
  Phys. Rev.} {\bfseries D78} (2008) 106003},
\href{http://arxiv.org/abs/0803.3085}{{\ttfamily arXiv:0803.3085 [hep-th]}}.

\bibitem{McAllister:2008hb}
L.~McAllister, E.~Silverstein, and A.~Westphal, ``{Gravity Waves and Linear
  Inflation from Axion Monodromy},''
  \href{http://dx.doi.org/10.1103/PhysRevD.82.046003}{{\em Phys. Rev.}
  {\bfseries D82} (2010) 046003},
\href{http://arxiv.org/abs/0808.0706}{{\ttfamily arXiv:0808.0706 [hep-th]}}.

\bibitem{Berg:2009tg}
M.~Berg, E.~Pajer, and S.~Sjors, ``{Dante's Inferno},''
  \href{http://dx.doi.org/10.1103/PhysRevD.81.103535}{{\em Phys. Rev.}
  {\bfseries D81} (2010) 103535},
\href{http://arxiv.org/abs/0912.1341}{{\ttfamily arXiv:0912.1341 [hep-th]}}.

\bibitem{Kappl:2014lra}
R.~Kappl, S.~Krippendorf, and H.~P. Nilles, ``{Aligned Natural Inflation:
  Monodromies of two Axions},''
  \href{http://dx.doi.org/10.1016/j.physletb.2014.08.045}{{\em Phys. Lett.}
  {\bfseries B737} (2014) 124--128},
\href{http://arxiv.org/abs/1404.7127}{{\ttfamily arXiv:1404.7127 [hep-th]}}.

\bibitem{Ben-Dayan:2014zsa}
I.~Ben-Dayan, F.~G. Pedro, and A.~Westphal, ``{Hierarchical Axion Inflation},''
  \href{http://dx.doi.org/10.1103/PhysRevLett.113.261301}{{\em Phys. Rev.
  Lett.} {\bfseries 113} (2014) 261301},
\href{http://arxiv.org/abs/1404.7773}{{\ttfamily arXiv:1404.7773 [hep-th]}}.

\bibitem{Shiu:2013wxa}
G.~Shiu, P.~Soler, and F.~Ye, ``{Milli-Charged Dark Matter in Quantum Gravity
  and String Theory},''
  \href{http://dx.doi.org/10.1103/PhysRevLett.110.241304}{{\em Phys. Rev.
  Lett.} {\bfseries 110} no.~24, (2013) 241304},
\href{http://arxiv.org/abs/1302.5471}{{\ttfamily arXiv:1302.5471 [hep-th]}}.

\bibitem{McAllister:2016vzi}
L.~McAllister, P.~Schwaller, G.~Servant, J.~Stout, and A.~Westphal, ``{Runaway
  Relaxion Monodromy},''
\href{http://arxiv.org/abs/1610.05320}{{\ttfamily arXiv:1610.05320 [hep-th]}}.

\bibitem{Hebecker:2017lxm}
A.~Hebecker, P.~Henkenjohann, and L.~T. Witkowski, ``{Flat Monodromies and a
  Moduli Space Size Conjecture},''
  \href{http://dx.doi.org/10.1007/JHEP12(2017)033}{{\em JHEP} {\bfseries 12}
  (2017) 033},
\href{http://arxiv.org/abs/1708.06761}{{\ttfamily arXiv:1708.06761 [hep-th]}}.

\bibitem{delaFuente:2014aca}
A.~de~la Fuente, P.~Saraswat, and R.~Sundrum, ``{Natural Inflation and Quantum
  Gravity},'' \href{http://dx.doi.org/10.1103/PhysRevLett.114.151303}{{\em
  Phys. Rev. Lett.} {\bfseries 114} no.~15, (2015) 151303},
\href{http://arxiv.org/abs/1412.3457}{{\ttfamily arXiv:1412.3457 [hep-th]}}.

\bibitem{Fisher:2017dbc}
Z.~Fisher and C.~J. Mogni, ``{A Semiclassical, Entropic Proof of a Weak Gravity
  Conjecture},''
\href{http://arxiv.org/abs/1706.08257}{{\ttfamily arXiv:1706.08257 [hep-th]}}.

\bibitem{Casini:2008cr}
H.~Casini, ``{Relative entropy and the Bekenstein bound},''
  \href{http://dx.doi.org/10.1088/0264-9381/25/20/205021}{{\em Class. Quant.
  Grav.} {\bfseries 25} (2008) 205021},
\href{http://arxiv.org/abs/0804.2182}{{\ttfamily arXiv:0804.2182 [hep-th]}}.

\bibitem{Preskill:1990ty}
J.~Preskill, ``{Quantum hair},''
\href{http://dx.doi.org/10.1088/0031-8949/1991/T36/028}{{\em Phys. Scripta}
  {\bfseries T36} (1991) 258--264}.

\end{thebibliography}\endgroup
\bibliographystyle{utphys}

\newpage
\section*{Erratum}

The convention for the gauge coupling
introduced in Sec.~\ref{subsec:verify}
was appropriate
when all the charged particles known 
to the low energy observers have
integer multiples of the smallest positive charge known.
For more general cases, I use
the convention for the gauge coupling 
described below.

I first define the unit charge
as the largest positive charge 
with which
all the charges of the particles 
known to the low energy observers
become integers.
Here, I assumed that the charges are quantized and this is always possible.
I also assume that
if a charged particle is known to the low energy observers,
they take it for granted that there exists 
corresponding anti-particle with the opposite charge
due to the $CPT$ theorem.
Then, since for every known charge $q$, 
there exists the opposite charge $-q$,
whether a charge is positive or negative is a matter of convention.
Then, as would be expected from the nomenclature,
I use the normalization convention of the gauge coupling 
in which the unit charge is one.

The normalization convention of the gauge coupling I chose
is convenient for describing the
Dirac quantization condition.
Indeed, with the above convention,
it is an elementary exercise in 
number theory to show
that there exists a combination
of charged particles known to the low energy observers
whose total charge is one.
Then, the condition that
the Dirac string should not be observable
by the Aharonov-Bohm phase of
such combination of charged particles
going around the Dirac string
gives 
the Dirac quantization condition:
\begin{equation}
q_m =  n \, ,
\label{eq:DiracQ2}
\end{equation}
where $q_m$ is the magnetic charge of the monopole
with the unit of the magnetic charge being 
$g_m = {2\pi}/{g}$,
and $n$ is some integer.

With my convention of the gauge coupling $g$ described above, 
the bound on the UV cut-off scale 
following from the magnetic WGC is given as
\begin{equation}
\Lambda \lesssim g M_P \, .
\label{eq:cutWGC2}
\end{equation}

It is straightforward to extend
the proof given in Sec.~\ref{subsec:trivial}
that the electric WGC is automatically satisfied
when there exists a charged particle
whose mass is below the UV cut-off scale
\eqref{eq:cutWGC} 
to more general cases discussed above.
In more general cases,
the UV cut-off scale following from the magnetic WGC
is given by \eqref{eq:cutWGC2},
with the convention of the gauge coupling $g$ explained above.
The proof can be given 
following the similar line of reasoning given in
\eqref{eq:light},
and noticing that
all the absolute value of charges
are greater or equal to one
in the current convention of the gauge coupling.

\end{document}